\documentclass{aa}

\newcommand{\orcid}[1]{\protect\href{https://orcid.org/#1}{\protect\includegraphics[width=8pt]{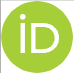}}}

\usepackage{hyperref}
\usepackage{longtable}
\usepackage{float}
\usepackage{placeins}
\usepackage{lineno}
\usepackage{amsmath,bm}
\graphicspath{{./}{figures/}}
\usepackage{graphicx}
\usepackage{placeins}
\usepackage{mathtools, nccmath}
\usepackage{lipsum}
\usepackage{siunitx,etoolbox}
\usepackage{capt-of}
\usepackage{float}
\usepackage[graphicx]{realboxes}
\usepackage[abs]{overpic}
\usepackage{mathrsfs}
\usepackage{txfonts}

\begin{document}

    \title{ReveaLLAGN 2: Exploring the Nuclear Gas and Stellar Kinematics of Nearby LLAGN with JWST/NIRSpec IFU}   
\titlerunning{ReveaLLAGN 2: Gas and Stellar Kinematics of Nearby LLAGN with JWST/NIRSpec IFU}
    
    \author{ Antoine Dumont \inst{1} \orcid{https://orcid.org/0000-0003-0234-3376}, Nadine Neumayer \inst{1} \orcid{https://orcid.org/0000-0002-6922-2598},  Anil C. Seth \inst{2} \orcid{0000-0003-0248-5470},  Torsten Böker \inst{6} \orcid{0000-0002-5666-7782}, Jonelle L. Walsh \inst{3} \orcid{0000-0002-1881-5908}, Michael Eracleous \inst{7} \orcid{0000-0002-3719-940X }, Luis C. Ho \inst{4,5} \orcid{0000-0001-6947-5846},Kameron Goold \inst{2} \orcid{0000-0002-7743-9906}, David Ohlson \inst{2} \orcid{https://orcid.org/0009-0004-9457-2495} \and Timothy A. Davis \inst{8} \orcid{https://orcid.org/0000-0003-4932-9379} }
    \authorrunning{Dumont et al. }

    \institute{Max Planck Institute for Astronomy, K{\"o}nigstuhl 17, 69117 Heidelberg, Germany\\ \email{antoine.dumont.neira@gmail.com} \and Department of Physics and Astronomy, University of Utah\\ 115 South 1400 East, Salt Lake City, UT 84112, USA \and  George P. and Cynthia Woods Mitchell Institute for Fundamental Physics and Astronomy and Department of Physics and Astronomy, Texas A\&M University, College Station, TX 77843, USA \and  Kavli Institute for Astronomy and Astrophysics, Peking University, Beijing 100871, China \and Department of Astronomy, School of Physics, Peking University, Beijing 100871, China \and European Space Agency, Space Telescope Science Institute, Baltimore, Maryland, USA \and Department of Astronomy \& Astrophysics and Institute for Gravitation and the Cosmos, The Pennsylvania State University, 525 Davey Lab, University Park, PA 16802 \and Cardiff Hub for Astrophysics Research \&\ Technology, School of Physics \&\ Astronomy, Cardiff University, Queens Buildings, Cardiff, CF24 3AA, UK}

    \abstract
    {{\it Context:} Low-luminosity active galactic nuclei (LLAGN) represent the most common mode of black hole accretion in the local Universe, yet their nuclear structure, stellar and gas kinematics remain poorly constrained at parsec scales.

    {\it Aims:} We investigate the nuclear regions of eight nearby representative LLAGN, the ReveaLLAGN sample: M87, M104, M94, M81, NGC~1052, NGC~4258, NGC~5128 and NGC~4395. Using the high spatial resolution of the Near Infrared Spectrograph (NIRSpec) integral field unit on the James Webb Space Telescope (JWST), we probe the stellar, molecular hydrogen, hydrogen recombination, and high ionization coronal lines kinematics and their spatial distribution, exploring how detectability and spatial distributions of these lines vary across different Eddington ratios.

    {\it Methods:} After cleaning  the JWST/NIRSpec data of point-spread-function sinusoidal artifacts, we spatially binned the data to increase its signal-to-noise. We fit the binned spectra using {\sc pPXF} with stellar templates and several gas emission lines, including H$_2$ molecular lines, H$_2$ recombination lines (Pa$\alpha$ and Br$\gamma$), and coronal lines such as [Si VI], [Mg IV], and [Ar VI], adopting a Gaussian profile to derive their kinematics. 

    {\it Results:} The stellar line-of-sight velocity maps reveal a slowly rotating disk in the nuclei of all our targets, characterized by an anti-correlation with the h$_3$ Gauss–Hermite moment. The stellar velocity dispersion rises toward the central black hole. Only in M104 and M94 does the velocity dispersion display a more intricate pattern that departs from a simple disk configuration, indicating the presence of multiple kinematic components in these galaxies. Across the sample, the gas distribution and kinematics exhibit rotation that is frequently decoupled from the stellar rotation, and often visually aligned with the radio jet, presenting broader velocity dispersions along the jet direction. Line-ratio maps of H$_2$ 1–0 S(1) to Br$\gamma$ indicate that the dominant excitation mechanism is photoionization by the central AGN, with mechanical heating from shocks becoming important only at higher Eddington rations.

    {\it Conclusions:} Across the ReveaLLAGN sample, we observe slow, disk-like stellar rotation that is decoupled from the gas kinematics, suggesting that the stellar gravitational potential is largely decoupled from the current accretion state. The gas kinematics are more diverse across the targets, with several sources showing kinematic features spatially associated with the radio jets. Across the sample, we detect coronal lines in all LLAGN when using nuclear-aperture spectra. However, no major dependence on Eddington ratio is found, except in the brightness of the coronal lines, which are sufficiently bright to obtain spatially resolved kinematics only in the highest-Eddington-ratio sources, NGC~5128 and NGC~4395.
    }

    \maketitle
\section{Introduction}
\label{sec:intro}

The majority of supermassive black holes (SMBHs) accrete material from their surrounding accretion disks at rates far below the theoretical Eddington limit \citep{Hasinger2005,Ho2008}. This is commonly expressed in terms of the Eddington ratio, defined as the ratio between their bolometric luminosity (L$_{\rm bol}$) and their Eddington luminosity (L$_{\rm Edd}$). Such systems are referred to as low-luminosity active galactic nuclei (LLAGN) and are distinguished by Eddington ratios of $\lesssim 0.01$. Although LLAGN are far more numerous than their high-Eddington ratio counterparts {\citep{Hasinger2005,Ho2008}, they remain relatively poorly understood because their intrinsic emission is weak, making it challenging to disentangle the nuclear contribution from the host galaxy light.

In contrast to quasars and Seyfert galaxies, LLAGN may also exhibit different accretion flows: the gas density is reduced, making the inner accretion disk optically thin and geometrically thick, so that the inflowing gas radiates inefficiently and forms a radiatively inefficient accretion flow (RIAF) \citep[e.g.][]{Narayan1995,Quataert2001}. As a result, LLAGN tend to show a diminished or absent broad-line region \citep[e.g.][]{Elitzur2009,Elitzur2014}, a truncated or absent accretion disk \citep[e.g.][]{Yuan2009,Eracleous2010,Ronald2012,Nemmen2014,Ontiveros2023}, and lacking the “blue bump” seen in brighter AGN\citep{Ho1999,Yuan2014}. Additionally, LLAGN observations also reveal a similar correlation between X-ray emission from the hot corona and the infrared emission from dust reprocessing, as observed in brighter AGN \citep{Kotilainen1992,Alonso-Herrero1996,Civano2012, Burtscher2015,Muller-Sanchez2018,Ohlson2026}, suggesting a possible scaled down AGN scenario for LLAGN. But deviations from this correlation have been found at extremely low-luminosities \citep{Dumont2020}.

The mass of the central SMBH is a fundamental parameter to characterize the AGN. Dynamical measurements of black holes masses in nearby galaxies have revealed a correlation between the SMBH mass (M$_{{\rm BH}}$) and the stellar velocity dispersion of the host bulge ($\sigma_{{\rm e}}$), known as the M$_{\rm BH}$–$\sigma_{\rm e}$ relation \citep[e.g.][]{McConell2012,Kormendy2013,Saglia2016}. This key relationship is based on both gas and stellar dynamical measurements. The consistency of these methods has been tested only in a handful of cases \citep[e.g][]{Neumayer2007,Cappellari2009,Feldmeier2014,Feldmeier2017,Krajnovic2018,Davis2020}, and the measurements do not always agree \citep[e.g][]{Ferrarese1996,Francesco2006,Walsh2012,Walsh2013,Gebhardt2011,Thater2021}. Since M$_{{\rm BH}}$ is small compared to the galaxy bulge mass ($\sim 0.3\%$) \citep{McConell2012} gravitational influence cannot explain the M$_{\rm BH}$–$\sigma_{\rm e}$ relation, and instead radiation pressure from AGN feedback is thought to be responsible. This relation suggests that SMBHs and their host galaxies may have grown in lockstep, with SMBHs potentially playing a significant role in galaxy formation \citep[e.g.][]{Croton2006}.

AGN feedback can be radiative, or be due to shocks or outflows from a jet.  This feedback can be traced through its impacts on the gas surrounding the AGN, disturbing its kinematics and changing its ionization state. The gas and jet interaction becomes particularly important at low accretion rates, because LLAGNs show a possibly enhanced jet activity \citep[e.g.][]{Ho2002,Sikora2007,Ho2008}. Jets can accelerate and compress the gas and produce shocks. These process can produce observable non-circular signatures in the gas kinematics \citep[e.g][]{Riffel2008,Davies2009,Hicks2009,Shimizu2019,Riffel2020}. Jets have also been observed to be responsible for producing extended coronal emission in some LLAGNs \citep{May2018,Rodriguez2025}. In the ReveaLLAGN sample, \citet{Kameron2024} suggested that outflows may explain the increase in emission line broadening with ionization potential observed in some LLAGNs. \citet{Kameron2025} found also that kinetic mode feedback may explain the higher excitation temperatures of molecular hydrogen in the ReveaLLAGN sample compared to brighter AGNs. At the same time, AGN feedback is closely linked to the availability of gas that can fuel SMBH accretion. Inflow of cold molecular gas in LLAGNs has been observed \citep[e.g][]{Combes2013} as well as inflow of warm molecular gas, for example in the LINER NGC~1097 \citep{Davies2009} and the Galactic Center \citep{Pei-Ying2017}. In other cases, such as NGC~5128, no signs of inflows are observed in the warm gas but there is a compact rotating disk \citep{Quillen2010,Espada2017}, indicating large reservoirs of gas in nuclear regions. Whether cold gas can reach the vicinity of the SMBH in LLAGNs, as observed in Seyfert galaxies, remains unclear. The enhanced radio loudness of LLAGNs compared with brighter AGNs may indicate that jets can expel gas before it reaches the innermost regions. One example is M87, whose jet has a very wide base, potentially consistent with this scenario \citep[e.g.][]{Chun2007}. 

To study the presence of inflows or gas interaction with the radio jet in the nuclear regions of LLAGN and to provide both new comparative gas and stellar dynamical masses we need high spatial resolutions kinematics.
The James Webb Space Telescope (JWST) integral field units (IFU) provide unmatched sensitivity and spatial resolution in the near- to mid-infrared range (1.7–28 $\mu$m$)$ \citep{NIRSpec_Boker, Wells2015,MIRI_MRS}, enabling detailed mapping of both stellar emission and ionized/molecular gas tracers in the nuclear regions of LLAGN.
Additionally, JWST's near-infrared (NIR) wavelength coverage provides access to high ionization potential lines. Compared with classical optical coronal lines, NIR coronal lines are less affected by dust obscuration and less contaminated by star formation from the host galaxy, making them distinct signatures of LLAGN accretion \citep{Satyapal2021}.

The primary objective of the ``Revealing Low-Luminosity Active Galactic Nuclei'' (ReveaLLAGN) project\footnote{\href{https://reveallagn.github.io/}{ReveaLLAGN webpage}} (Cycle 1 GO proposal, ID 2016, PI A. Seth) is to take advantage of JWST’s unparalleled sensitivity and angular resolution to investigate emission lines of nearby low‑luminosity AGN (LLAGN) and spectral energy distribution features that are unique to them. The ReveaLLAGN sample comprises eight prototypical systems with Eddington ratios from $\log(L_{bol}/L_{Edd})\approx-6$ to $\approx-2.7$ and distances of $\sim 4-20$ Mpc, corresponding to spatial resolutions of $\sim1-10$ pc, respectively (see Table~\ref{tab:targets_info} for details). Five of the ReveaLLAGN targets have Eddington ratios of $\leq 10^{-4.5}$, where a transition from a radiatively efficient disk to a more jet-dominated flow is expected \citep{Fender2003}. The proximity of these LLAGN allows us to study the stellar and gas kinematics of the nuclear regions at parsec-scale and resolve the sphere of influence of the SMBH which can provide crucial insights into their accretion mechanism. For the ReveaLLAGN sample we estimate an average black hole sphere of influence of $0.18\arcsec$ (for a distance of $\sim9$ Mpc, SMBH of $\sim10^{7.9}\,{\rm M_{\odot}}$ and velocity dispersion of $\sim200$ km s$^{-1}$) which is larger than the NIRSpec point spread function (PSF) full width half maximum (FWHM) of $0.15\arcsec$ \citep{Bentz2025}. The NIRSpec wavelength coverage allows us to trace simultaneously the stellar kinematics via the $^{12}$CO(2-0) absorption lines at $2.29\,\mu {\rm m}$ (hereafter CO bandhead), the warm gas through the H$_2$ transitions, ionized gas via hydrogen recombination lines and high-ionization coronal lines such as [Si VI], [Mg IV] and [Ar VI] at $1.963\,\mu {\rm m}$, $4.487\,\mu {\rm m}$ and $4.529\,\mu {\rm m}$ respectively. Coronal lines in the NIR originate from ions with high ionization potentials ($\geq \, 100$ eV) that require photoionization by compact and dense regions close to the SMBH \citep[e.g][]{Rodriguez-Ardilla2002,Muller-Sanchez2011}, potentially being very affected by dust obscuration \citep{Lamperti2017}. However, Mid-infrared coronal lines, such as [Ne V] ($14.3\,\mu {\rm m}$) and [O IV] ($25.9\,\mu {\rm m}$) have been identified in the ReveaLLAGN targets using JWST MIR/MRS data \citep{Kameron2024, Kameron2025}.

In this third paper of the ReveaLLAGN series, we present the JWST/NIRSpec IFU observations of seven objects in the ReveaLLAGN sample: M104, M81, M87, M94, NGC~1052, NGC~4258, NGC~4395, as well as the GTO data for the nuclear region of NGC~5128. We explore the central $3\arcsec \times3 \arcsec$ taking advantage of the high-spatial ($0.1\arcsec$) and spectral resolution (${\rm R}\sim2700$) of the NIRSpec IFU to get resolved maps of stellar, molecular, and ionized gas kinematics, as well as performing a search for NIR coronal line emission. 
The kinematics presented here will be used in a future work involving the careful dynamical modeling of the gas in the nuclear regions of this sample. However, these results can already be used to answer whether there is the presence of large gas reservoirs, orderly disk rotation and if there are clear visual perturbations that can be link to the jet.

Throughout this paper, all wavelengths are quoted as rest-frame vacuum wavelengths, unless otherwise noted. The paper is organized in five sections. Section~\ref{sec:DATA} discusses the data reduction as well as the removal of sinusoidal patterns with \texttt{WICKED} \citep{Dumont2025}. Section~\ref{sec:ReveaLLAGN_results} presents the stellar and gas kinematic results of the JWST/NIRSpec data of the ReveaLLAGN sample. In Section~\ref{sec:Summary} we give a summary and a discussion of our work. 

\section{Target Sample and Data}
\label{sec:DATA}
The galaxy sample studied in this paper consists of JWST nuclear observations of eight well-known LLAGN, listed in order of increasing Eddington ratio: M87, M104 (Sombrero), M94 (NGC~4736), M81, NGC~1052, NGC~4258 (M106), NGC~5128 (Cen~A), and NGC~4395. The sample covers a large range of black hole masses $10^{5.5-9.8}\, {\rm M}_{\odot}$ and Eddington ratio $\log{L_{bol}/L_{Edd}}$  of -6.2 to -2.7. The data for NGC~5128 come from the GTO program PID~1269 (PI: Nora Luetzgendorf) and the rest are from the Cycle 1 program PID~2016 (PI: Anil C. Seth). The data were taken using the high-resolution spectral modes G235H/F170LP and G395H/F290LP with a four-point dither pattern. In this work we present the NIRSpec IFU data of the ReveaLLAGN sample while the MIRI/MRS data are presented in a complementary paper, ReveaLLAGN 1 \citep{Kameron2025}. All data are publicly available on the ReveaLLAGN website \footnote{\url{https://reveallagn.github.io/}}. For the grating/filter combination used here, the IFU mode of NIRSpec obtains spectra in the wavelength range $0.6-5.3$ $\mu {\rm m}$ over a $3.1\arcsec \times 3.2\arcsec$ field of view at a spectral resolution of $R\approx2700$ \citep{NIRSpec_Boker} and with a PSF FWHM of $\sim0.15\arcsec$ \citep{Bentz2025}. 
The total integration time for the eight targets was 31.9 hours, with a typical on-source exposure time of $\sim 900$s per galaxy and per grating, to ensure a minimum signal-to-noise (S/N) of ${\rm S/N}\sim 30$. Table~\ref{tab:targets_info} shows the total integration time for the NIRSpec observations, as well as some relevant information for each target. The targets are sorted in order of increasing Eddington ratio ($\log{(L_{\mathrm{bol}}/L_{\mathrm{Edd}})}$). 

\begin{table*}
    \centering
    \begin{tabular}{|c|c|c|c|c|c|c|c|c|}
        \hline 
        Galaxy & Distance & 0.1\arcsec scale  & AGN type & M$_{\rm BH}$ & Eddington ratio & Radio Jet & Exp.& Source \\
         & (Mpc) & pc & &  $\log_{10}({\rm M_{\odot}})$ & $\log(L_{bol}/L_{Edd})$ &  & (s) & \\
        \hline 
        M87 & 15.1 & 8.0  & L2 & 9.8 & -6.0 & Yes & 2,050 & ReveaLLAGN\\
        M104 & 9.9 & 4.7 & L2 & 8.8  & -5.7  &  ? & 3,280 & ReveaLLAGN\\
        M94 & 4.2 & 2.4 & L2 & 6.8 & -5.2  & ? & 3,280 & ReveaLLAGN\\
        M81 & 3.6 & 1.7 & S1.5 & 7.8 & -4.5  & Yes & 3,280 & ReveaLLAGN\\
        NGC~1052 & 19.2 & 9.4 &  L1.9 & 8.8$^{*}$ & -4.2 & Yes & 2,050 & ReveaLLAGN \\
        NGC~4258 & 7.5 & 3.5 & S1.9 & 7.6 & -3.5 & Yes & 2,050 & ReveaLLAGN\\
        NGC~5128 & 3.4 & 1.7 & L2 & 7.7 & -2.7  & Yes & 1,167 & GTO\\
        NGC~4395 & 4.3  & 2.2 & S1.8 &  5.6 & -2.7 & Yes  & 2,050 & ReveaLLAGN\\
        \hline 
    \end{tabular}
    \caption{Target information for the NIRSpec IFU G235H/F170LP and G395H/F290LP observations in the ReveaLLAGN program, listed in order of increasing Eddington ratio. The second column shows the distance adopted from the literature: M87 \citep{Mendez1993}, M104 \citep{Sabbi2018}, M94 \citep{Radburn2011}, M81 \citep{McCommas2009}, NGC~1052 \citep{Tully2013}, NGC~4258 \citep{Riess2016}, NGC~5128 \citep{Ferrarese2007} and NGC~4395 \citep{Thim2004}. The third column shows the physical scale at the distance of the object for the NIRSpec high-res mode. \textbf{AGN type} shows the literature classification for the AGN source in each galaxy are taken from \citet{Ho2008}, and for NGC~5128 from \citet{Simpson1998A}. LINERS are denoted as L and Seyfert galaxies as S. The M$_{\rm BH}$ are taken from: M87 \citet{EHT2019}, M104 \citet{Jardel2011}, M94 \& M81 \citet{Kormendy2013}, NGC~1052 derived from scaling relations from \citet{Koss2022}, NGC~4258 \citet{Reid2019}, NGC~5128 \citet{Cappellari2009} and NGC~4395 \citet{Denbrok15}. \textbf{Eddington ratios} are from the ReveaLLAGN 1 paper \citet[(and references within]{Kameron2025}. \textbf{Radio Jet}: M87-\citet{Owen1989}, M104-\citet{Hada2013}, M94 \citet{Oers2017}, NGC~1052-\citet{Kadler2002}, NGC~4258 \citet{Herrnstein1997}, NGC~5128 \citet{Jones1996} and NGC~4395-\citet{Nandi2023}.
    \textbf{ Exp.} shows the on source integration time summed for both filters (G235H/F170LP and G395H/F290LP) in seconds.  }
    \label{tab:targets_info}
\end{table*}

The Eddington luminosities are computed from literature black hole masses (M$_{\rm BH}$) listed in Table~\ref{tab:targets_info} and are calculated as follow:
\begin{equation}
    {\rm L}_{Edd} = 1.26\times10^{38}\Bigg(\frac{\rm M_{BH}}{{\rm M_{\odot}}}\Bigg)\,{\rm erg\,s^{-1}}
\end{equation}
Bolometric luminosities were derived using scaling relations from \citet{Duras2020}, based on the nuclear 2-10 keV X-ray luminosities from \citet{Asmus2015}, except for M94, for which the 2-10 keV luminosity is taken from \citet{Gonzales2009}.

\subsection{Data Reduction}
\label{sec:data_reduction}
The NIRSpec IFU data were reduced using the JWST data pipeline version 1.12.5 and context file ``jwst$\_$1256.pmap'' 
\citep{Bushouse2023}. The data reduction was identical for the entire ReveaLLAGN sample, except for the saturated G395H/F290LP observations of NGC~5128; the modified procedure is described at the end of this subsection. We started the reduction by running the \texttt{Detector1} module of the JWST pipeline on all raw {\it uncal.fits} files. We used the patch \texttt{snowblind} \citep{snowblind} during \texttt{Detector1} to correct for large cosmic ray hits. For this,  we saved the ``jump.fits'' files during the \texttt{Detector1} and passed them to \texttt{snowblind}, which detected and masked pixels affected by cosmic-ray hits. 
The clean ``jump.fits'' files are passed again to \texttt{ Detector1} to run the remaining steps until obtaining a ``rate.fits'' file. The count-rate files were then processed using the \texttt{Calwebb\_spec2} module with default parameters, but using the \texttt{NSClean} \citep{NSClean} step (now incorporated into the JWST pipeline) to correct for correlated noise. Finally, the resulting ``cal.fits'' files were resampled and co-added using the ``drizzle'' weighting into a cube using the \texttt{Calwebb\_spec3} with a spaxel size of $0\farcs1$ and the instrument alignment.

The G395H/F290LP cube of NGC~5128 reduced in the fashion described above shows signs of saturation in spaxels where the AGN light dominates. After examining the individual {\it jump} files for each of the four dither images, all but the first dither image were saturated after the first group. Therefore, for NGC~5128 cube we run the \texttt{Rampfit} step in the \texttt{Detector1} using only the first integration group by setting \texttt{ detector1.ramp\_fit.suppress\_one\_group= False}, and excluding the rest of the dither images. This effectively reduced the integration time for the G395H/F290LP cube of NGC~5128 from 933s to 233s. 

\subsection{Wiggle correction}
\label{sec:WICKED_correction}
All data cubes present substantial sinusoidal modulations in the spectra at the single-spaxel level. This is a known consequence of PSF undersampling in NIRSpec \citep{Law2023}. These modulations, commonly referred to as ``wiggles'',  are particularly noticeable near the brightest spaxel, where the compact AGN light dominates.

The presence of these wiggles can significantly distort line profiles, which can affect the ability to obtain reliable kinematics. We used the \texttt{Python} tool \texttt{WICKED} \citep{Dumont2025} to remove the wiggles from the single-spaxel spectra. We found that the number of wiggle affected spaxels that we correct with \texttt{WICKED} range from eight spaxels (four in each filter) for M104 (our faintest galaxy) to $\sim 80$ spaxel (both filters combined) for some of our brightest galaxies like NGC~1052, NGC~4258 \& NGC~5128. The corrected spaxels are mostly located in the inner $0.5\arcsec$ of the cube near the bright point-like LLAGN. \citet{Dumont2025} tested the impact of removing the wiggles using \texttt{WICKED} and found that the corrected data showed a difference of less than $5\%$ in equivalent width across many emission lines and a difference of less than $1\%$ in the line-of-sight (LOS) velocity. A detailed description of the effectiveness of \texttt{WICKED} at removing wiggles is given in \citet{Dumont2025}. 

In Figure~\ref{fig:example_m94_spec} we illustrate the impact of not correcting for wiggles on the reliability of stellar and gas kinematics. It shows the {\sc pPXF} spectral fitting (performed as described in Section~\ref{sec:ReveaLLAGN_results}) of the brightest (central) Voronoi bin in M94, together with the derived kinematics along the Voronoi bins on the $45^\circ$ axis with respect to the cube $x$-axis ( $135^\circ$ east-of-north), approximately along the kinematic position angle (PA). The uncorrected spectra, shown in blue, clearly display strong sinusoidal modulations that heavily distort the spectra and compromise the ability of {\sc pPXF} to obtain a good fit, even when using a 10th-order additive and 6th-order multiplicative polynomial. For the uncorrected data, {\sc pPXF} overestimates both the stellar LOS velocity and dispersion; in M94, the two most central Voronoi bins show velocity dispersions that are at least $40\%$ higher. We also note that the kinematics reported by \citet{Nguyen25} agree well with those of our uncorrected data, likely leading to an overestimate of the central supermassive black hole mass. We note that the \texttt{WICKED} correction still leaves some residual wiggles in the spectrum, visible in the best-fit model around the first CO bandhead at $2.29,\mu$m, where there is a small mismatch between the line continuum and the model. This mismatch coincides with a wiggle at the same wavelength in the residuals that \texttt{WICKED} only partially removed. 

The wiggles also have a strong impact on the gas kinematics. Since the adjacent continuum of the emission lines is altered, {\sc pPXF} can either miss a line or mistakenly fit a wiggle as an emission line, as seen in Figure~\ref{fig:example_m94_spec} for the Br$\beta$ line at $2.62\,\mu$m. Therefore, careful correction of these artifacts is a crucial step in obtaining reliable stellar and gas kinematics. Finally, because the wiggles in neighboring spaxels are correlated, aperture extractions of radii roughly equal (or larger) to the PSF ($\sim0.15\arcsec$) will average them out from the spectrum \citep{Law2023}. When comparing aperture extracted spectra from uncorrected and \texttt{WICKED} cleaned cubes for every target in the ReveaLLAGN sample, we find an average flux difference of less than $3\%$, and their LOS velocities and velocity dispersion agreeing within their uncertainties. 

Recently, a new method for mitigating the wiggles in JWST/NIRSpec data was introduced in pipeline version 2.0.0 \citep{Law2026}. The method, known as the ``adaptive tracer model'' (ATM), was developed after the data reduction for the ReveaLLAGN project was completed. As a sanity check, we compared the performance of ATM and \texttt{WICKED} in removing the wiggles. We find an average difference in flux of 2$\%$ between the two methods, and the derived kinematics agree within the uncertainties. An average difference in flux of 5$\%$ was expected in \citet{Dumont2025} when correcting the wiggles using \texttt{WICKED}. We therefore find no major improvement when using ATM compared to \texttt{WICKED}, and we expect no significant biases in our analysis from using \texttt{WICKED} instead of the ATM method.

\begin{figure*}[h!]
\includegraphics[width=.49\textwidth]{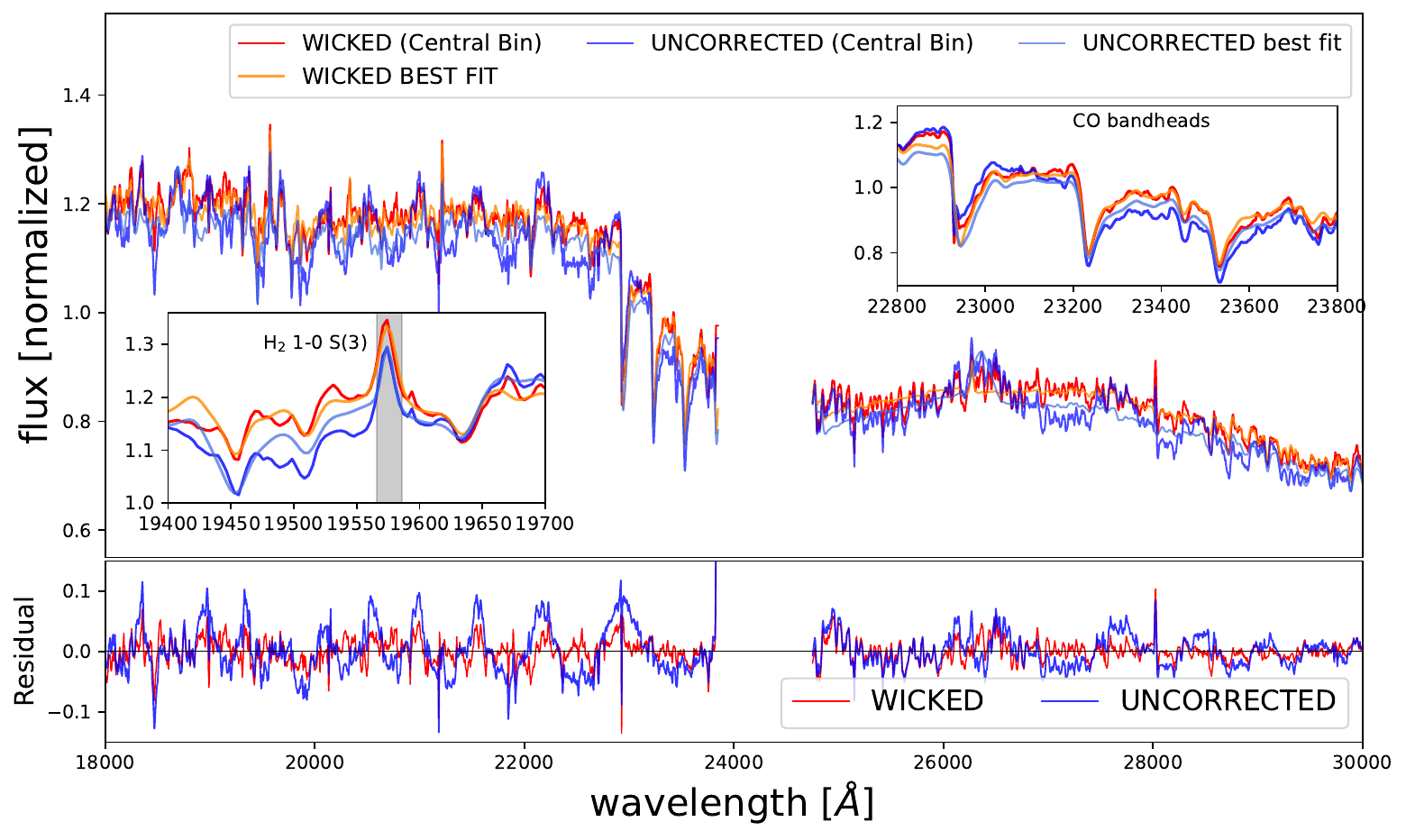}
\includegraphics[width=.49\textwidth]{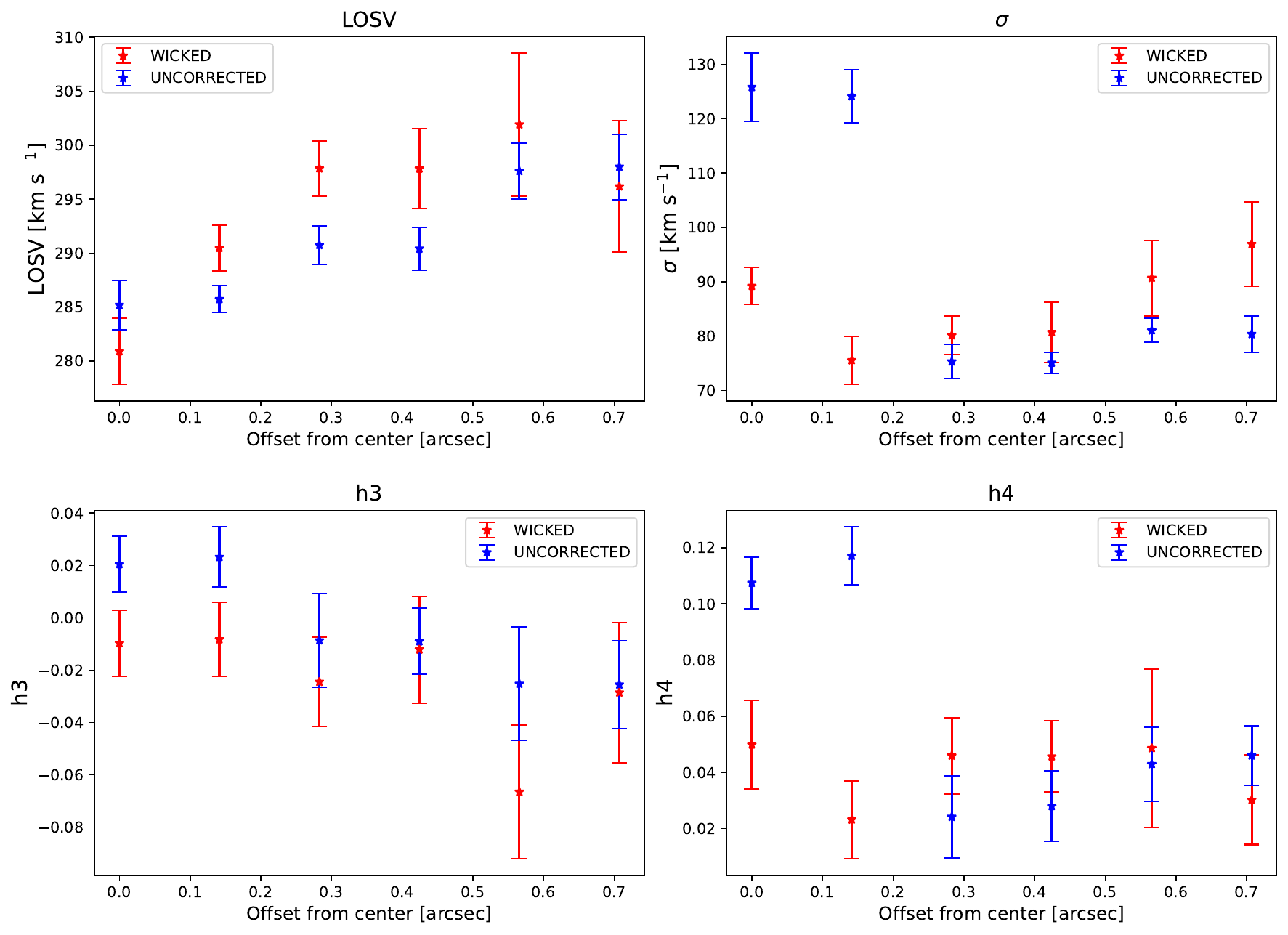}
\caption{{\it Left}: Brightest Voronoi-bin spectrum of M94 from the cube cleaned with \texttt{WICKED} (red) compared to the uncorrected cube (blue). The best {\sc pPXF} fit is shown in orange for the WICKED-cleaned cube and in blue for the uncorrected cube. Zoom-in panels highlight the H$_2$ 1-0 S(3) line at $1.95\,\mu$m and the stellar CO bandhead at $2.29\,\mu$m, illustrating how wiggles distort the line profiles and bias the derived kinematics. The bottom panel shows the corresponding residuals, clearly revealing wiggles in the uncorrected cube. {\it Right}:  Comparison of the stellar  LOS velocity (left) and velocity dispersion (right) for Voronoi bins along a $45^\circ$ axis for the WICKED-corrected and uncorrected cubes. In the central $\sim0.7\arcsec$, even in binned data, wiggles lead to an overestimation of the velocity dispersion in the inner $\sim0.15\arcsec$ and also affect the derived  LOS velocity. We note that this impact is even more significant for kinematics measured in individual spaxels.}
\label{fig:example_m94_spec}
\end{figure*}

\section{ReveaLLAGN: Stellar \& Gas Kinematics, Flux and Excitation Maps}
\label{sec:ReveaLLAGN_results}
In this section, we discuss the stellar and gas kinematics maps. We obtain flux maps for both the warm molecular and ionized hydrogen components (Pa$\alpha$, Br$\gamma$) and a line ratio map between the H$_2$ 1-0 S(1) and Br$\gamma$ to investigate the dominant excitation mechanism of the warm molecular gas. These nuclear kinematic data will be used to measure dynamical black hole masses in an upcoming paper \textit{Dumont et al (in prep.)}.  

Figure \ref{fig:example_spec_ReveaLLAGN} shows the nuclear spectra for the entire sample for an aperture extraction of $0.3\arcsec$ (roughly twice the size of the PSF). We recognize many gas emission lines in the nuclear spectrum of most galaxies as well as the common CO stellar absorption feature at $2.29\,\mu {\rm m}$ (vertical blue lines). This stellar feature is a commonly used tracer of stellar kinematics in the NIR in galaxy centers \citep[e.g.][]{Cappellari2009,Riffel2009,Seth2010m32,Denbrok15,Fahrion2024,AlAmri2025}. Based on the available line list from the Gemini Observatory\footnote{https://www.gemini.edu/observing/resources/near-ir-resources/spectroscopy/important-h2-lines}, we recognize three important ro-vibrational hydrogen molecular lines  (shown in yellow vertical lines in Figure~\ref{fig:example_spec_ReveaLLAGN}), 1-0 S(3), 1-0 S(2), 1-0 S(1), 1-0 O(5) and 0-0 O(9) at 1.9576 $\mu{\rm m}$, 2.0338 $\mu {\rm m}$, 2.1218 $\mu {\rm m}$, 3.2350 $\mu {\rm m}$ and 4.6947 $\mu {\rm m}$ respectively. These lines are a great tracer of warm molecular gas in LLAGN \citep[e.g][]{Riffel2013}. Using the same line list,  we also identify a series of prominent hydrogen recombination lines, such as Pa${\alpha}$ $1.875\,\mu {\rm m}$, Br${\gamma}$ $2.166\,\mu {\rm m}$ and Br${\beta}$ $2.625\,\mu {\rm m}$, as well as three of the Hydrogen Pfund series Pf${\delta}$ $3.296\,\mu {\rm m}$, Pf${\gamma}$ $3.740\,\mu {\rm m}$ and Pf${\beta}$ at $4.653\,\mu {\rm m}$ all with an ionization potential of $\sim 14$ eV, which is why they can be present in star forming regions and AGN. We also detect various NIR coronal lines in the nuclear spectra. In the G235H/F170LP part of the spectrum we detect [Mg VIII] and [Fe IV] tracing ionization potentials $\sim 50$ eV at 3.0280 $\mu {\rm m}$ and 2.8050 $\mu {\rm m}$. We also detect [Si VI] tracing ionization potentials of 167 eV at 19630 $\mu {\rm m}$, making it the highest excitation potential line that we detect. In the G395H/F290LP, we detect [Mg IV] $4.487\,\mu {\rm m}$ and [Ar VI] $4.530\,\mu {\rm m}$, with ionization potentials of $97-124$ eV. These coronal lines arise from regions within a few parsecs of the central SMBH and are considered unambiguous signatures of AGN photoionization \citep[e.g]{Genzel1998,Lutz1999,Satyapal2007,Satyapal2008,Satyapal2009,Goulding2009,Veilleux2009,Satyapal2021}.

\begin{figure*}[h!]
\includegraphics[width=.99\textwidth]{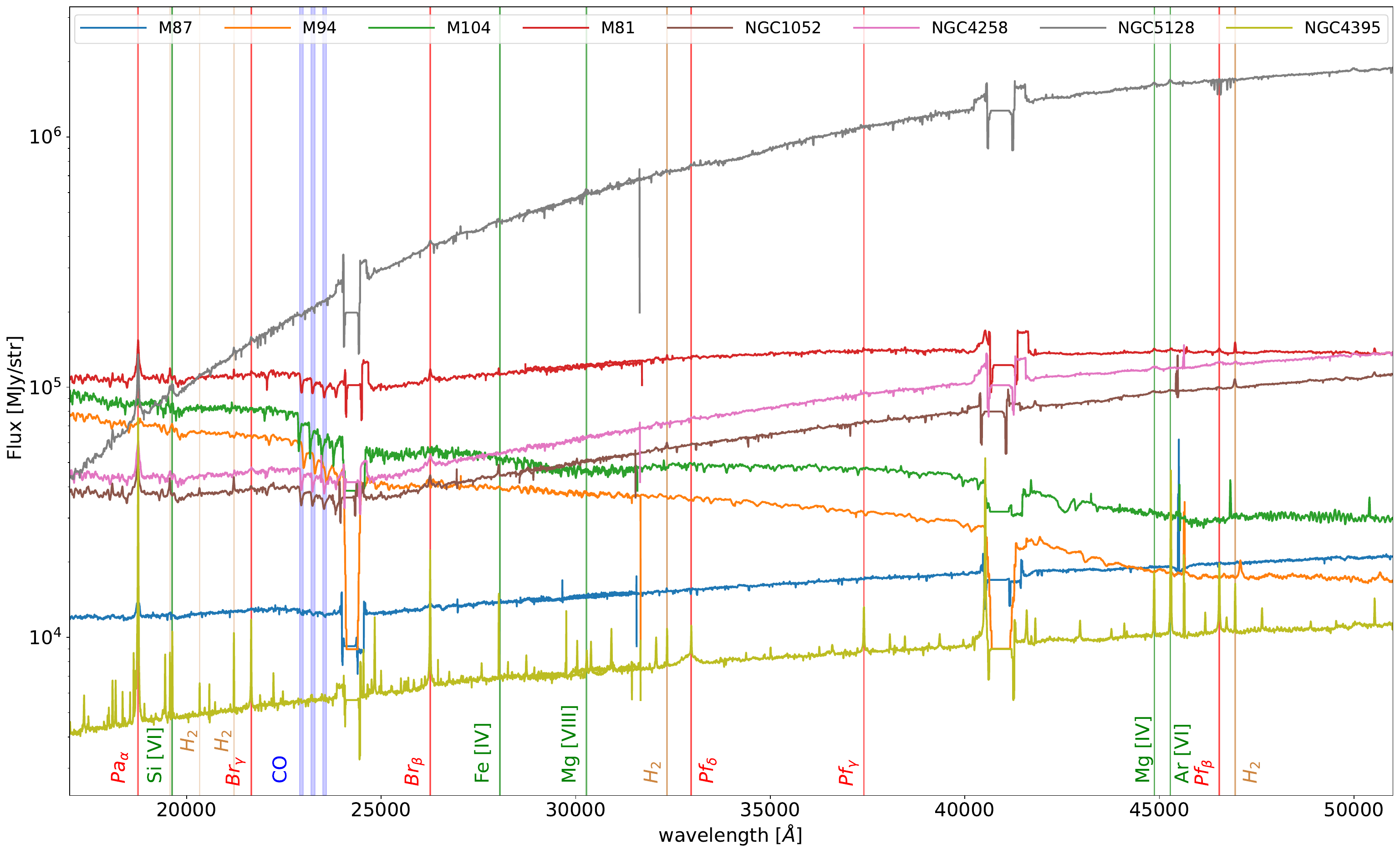}
\caption{ The $0.3\arcsec$ nuclear aperture spectrum for the full ReveaLLAGN sample and NGC~5128. Vertical lines indicate different gas emission lines that we fit. Yellow lines show the molecular hydrogen lines, red lines are the ionized hydrogen lines, and green lines denote the coronal lines. We also highlight the $^{12}$CO(2-0) lines at $2.9\,\mu {\rm m}$, the strongest stellar feature present at these wavelengths. }
\label{fig:example_spec_ReveaLLAGN}
\end{figure*}

In the rest of this section we provide a general overview of the method for deriving the stellar kinematics (S~\ref{subsec:stellar_kin}), gas kinematics (\S~\ref{subsec:gas_kin}), and flux and excitation maps (\S~\ref{subsec:flux_distribution}). In subsequent subsections, we provide a more in-depth discussion for each galaxy.

\subsection{Stellar kinematics fits}
\label{subsec:stellar_kin}

For our kinematic fits, we spatially binned the data cubes using the {\sc Voronoi python package} \citep{Cappellari2003} to increase the S/N of the spectrum. We could not define a universal S/N target due to the large diversity in continuum luminosity for our sample, so instead we aimed to find a S/N target that produced $\sim 500$ spatial bins. With this number of bins, the inner $\sim 0.5\arcsec$ typically have single spaxel spectra, ensuring high spatial resolution to resolve the central kinematics. This resulted in a continuum S/N of $60-100$ for most of our sample, except for NGC~4395, for which the continuum S/N is $25$, while NGC~4395 still exhibits significantly strong emission lines.

We used the {\sc Python} implementation of {\sc pPXF} \citep{Cappellari2017} to fit
the LOS velocity (v), the velocity dispersion ($\sigma$), and the Gauss-Hermite velocity moments corresponding to skewness (h$_3$) and kurtosis (h$_4$). The fit was done with a set of synthetic high-resolution stellar templates (R$\sim 100000$) from the Phoenix library \citep{phoenixlib}. We selected templates based on Padova isochrones \citep{Padova-iso} of 10 Gyr with metallicities ranging from Z$_{\odot} = 0.002$ to Z$_{\odot}= 0.056$, to primarily sample cool luminous stars, which dominate the K-band light and present CO bandhead absorption features commonly used to obtain stellar kinematics at these wavelengths \citep[e.g.,][]{Silge2003,Cappellari2009}.
This led to a selection of 57 templates with effective temperatures (T$_{{\rm eff}}$) ranging from 3000-6700 K, metallicities of $-2\leq {\rm [Fe/H]} \leq 1$ and surface gravity $\log(g)=0-4$. We convolved the stellar templates to match the instrumental spectral resolution of the NIRSpec G235H/F170LP and G395H/F290LP (R$\sim 2700$).

We tested the robustness of the derived stellar kinematics by varying the degree of the additive and multiplicative polynomials used in the {\sc pPXF} fits. The stellar LOS velocity, dispersion, and higher-order Gauss-Hermite moments converged after using a multiplicative and additive polynomial of n $=6$ and n $=10$, respectively. Based on this, we adopted these polynomials for our fits, which are also similar to the ones used for stellar kinematics using NIRSpec data in previous works \citep[e.g][]{Fahrion2024,AlAmri2025}. The 10th-degree additive polynomial is required to model the varying continuum in the spectrum coming from AGN activity \citep[e.g][]{Fahrion2024}. The inclusion of the 6th-degree multiplicative polynomial reduced the stellar velocity dispersion by an average of $\sim50$ km s$^{-1}$, which reconciles the larger-than-expected velocity dispersion in the inner $\approx0.3\arcsec$ previously reported for NGC~5128 in \citet{Dumont2025} using the same NIRSpec G235H/F170LP data with previous ground-based data presented in \citet{Cappellari2009}. 

Although we could fit the full spectrum for both JWST filters to constrain the stellar kinematic, in practice the G395H/F290LP spectra lack strong stellar absorption features. Therefore, we restricted our stellar fits to the G235H/F170LP data, which cover the CO bandhead stellar absorption lines at $2.29\,\mu$m which is the most notorious stellar feature at these wavelengths \citep[e.g.,][]{Silge2003,Cappellari2009}. We restricted the stellar fits to the wavelength range between $1.7-3.16\,\mu {\rm m}$ which excludes the spectrum edges and the detector gap between $2.380-2.475\,\mu {\rm m}$. An example of our pPXF fits is shown with the orange line in Fig.~\ref{fig:example_m94_spec}.

Figure~\ref{fig:stellar_kinematic_maps} shows the stellar kinematic maps of the first four Gauss-Hermite velocity moments for the LLAGN sample, excluding M87 and NGC~4395. NGC~4395 does not show any stellar absorption lines in any of the filters, either at single-spaxel level or in large aperture spectra; therefore, we cannot constrain its stellar kinematics. Similarly, the M87 spectra show very weak stellar features and were heavily affected by wiggles in the nuclear regions, preventing us from deriving stellar kinematics. 

\begin{figure*}
\includegraphics[width=.95\textwidth]{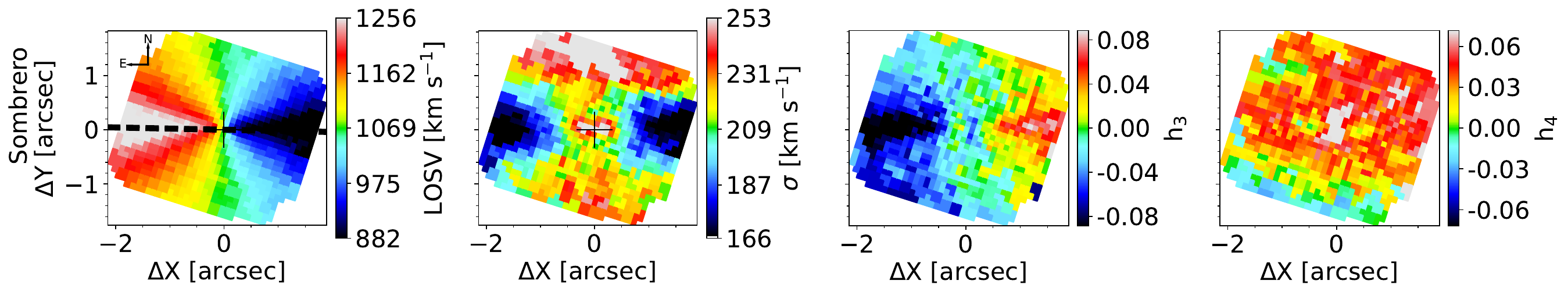}
\includegraphics[width=.95\textwidth]{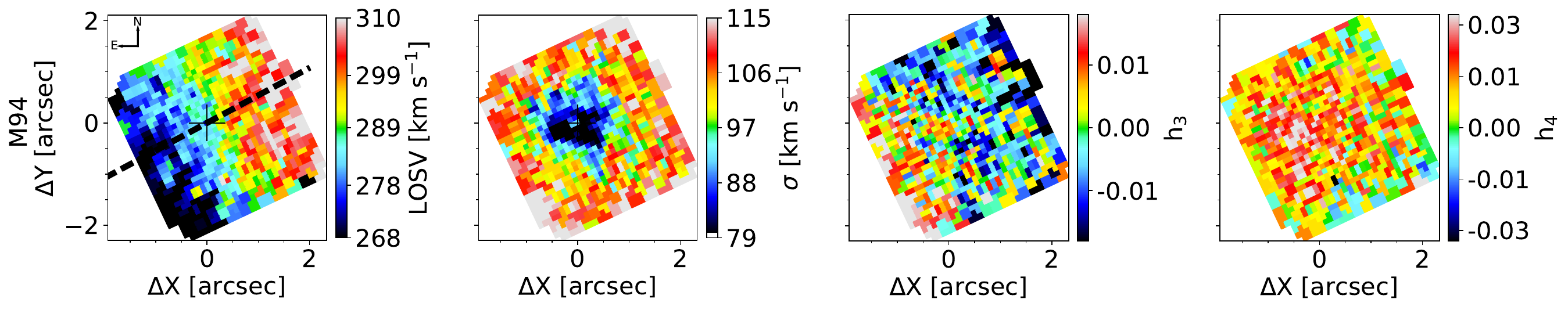}
\includegraphics[width=.95\textwidth]{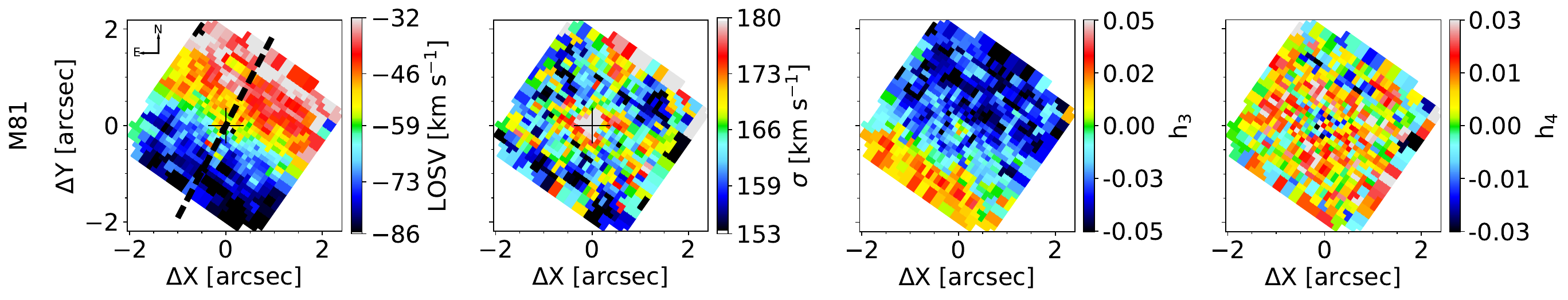}
\includegraphics[width=.95\textwidth]{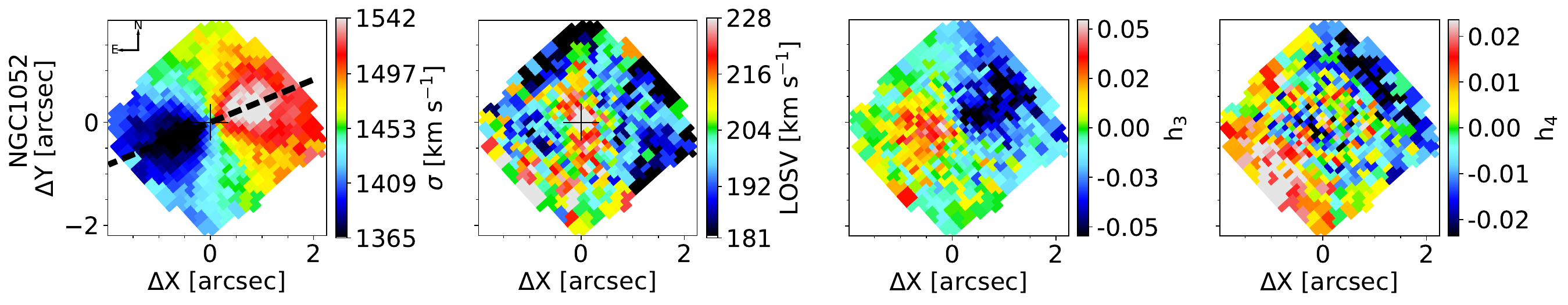}
\includegraphics[width=.95\textwidth]{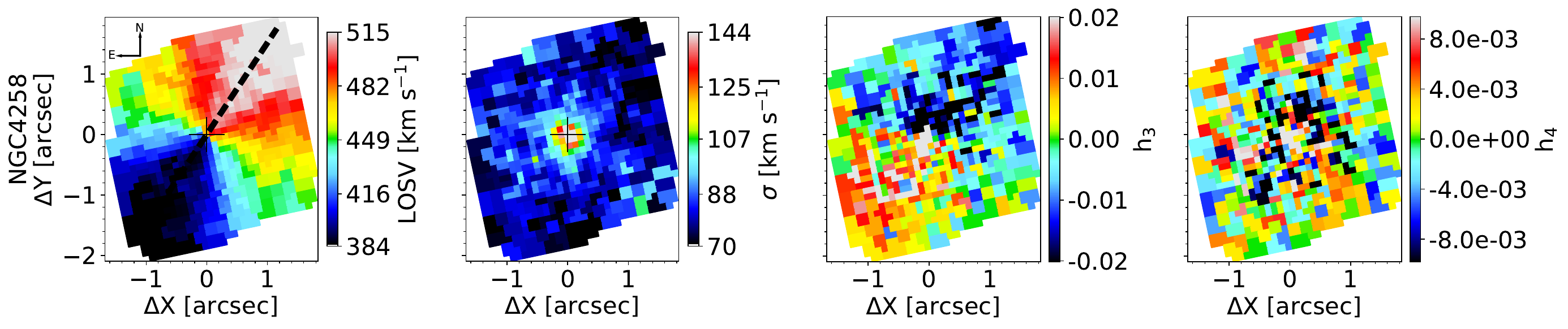}
\includegraphics[width=.95\textwidth]{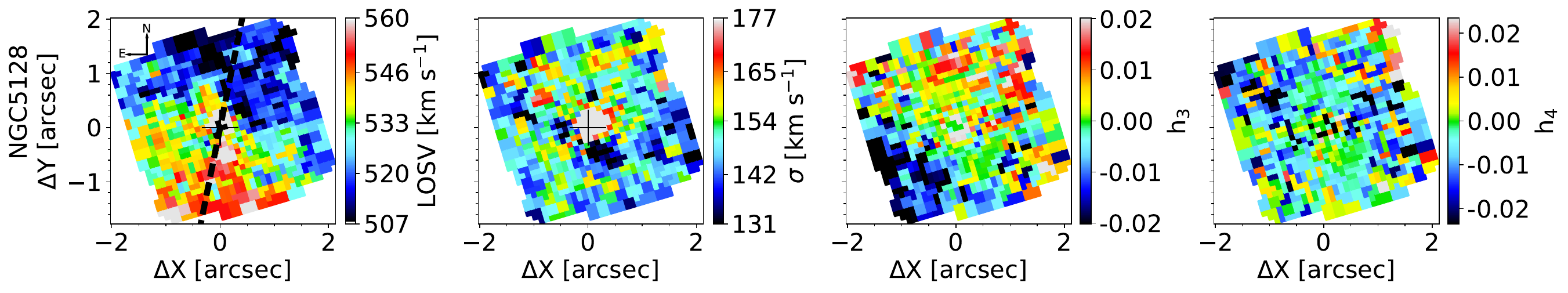}

\caption{ReveaLLAGN stellar kinematic maps for the first four velocity moments: velocity, velocity dispersion, h$_3$, and h$_4$. The maps are oriented north up and east to the left. The dashed black line shows the best fit stellar kinematic PA. }
\label{fig:stellar_kinematic_maps}
\end{figure*}

\textbf{Stellar LOS velocities}. The LOS velocity maps presented in the left column of Figure~\ref{fig:stellar_kinematic_maps} show clear signatures of regular disk-like rotation for all galaxies. For each object, we derive a global kinematic PA using the {\sc Python} package {\sc PaFit} \citep{Krajnovic2006}. The kinematic PA traces the rotation axis, along which the projected LOS velocity gradient is maximum, and provides an estimate of the rotation amplitude $\Delta$V. Three of our galaxies, M104, NGC~1052, and NGC~4258, which are among the highly inclined system in our sample, show stronger rotation than the rest of the sample, with rotation amplitudes of $\Delta V_{LOSV} \approx 320$ km s$^{-1}$, $\approx 180$ km s$^{-1}$ and $\approx 120$ km s$^{-1}$, respectively. M94 and NGC~5128 show much more modest rotation with $\Delta V_{LOSV} \leq 50$ km s$^{-1}$. Finally, for M87 and NGC~4395 we could not derive stellar kinematics because of their low-surface brightness cores and the impact of wiggles in spectra from AGN activity in the nuclear regions. 

\textbf{Stellar velocity dispersion}. 
The second column in Figure~\ref{fig:stellar_kinematic_maps} shows the velocity dispersion maps of the LLAGN sample. We observe a rise in the velocity dispersion towards the center for all objects except for M94, where a drop is observed instead. A rise in the velocity dispersion at the center is expected for regions dominated by the gravitational potential of a supermassive black hole, but stellar structures on smaller scales or a lack of stars near the black hole can create exceptions. For M94 the stellar dispersion drop at the center appears to be associated with a counter-rotating structure at the center. 

\textbf{Higher Gauss-Hermite velocity moments}. 
The two right panels in Figure~\ref{fig:stellar_kinematic_maps} show the Gauss-Hermite velocity moments h$_3$ and h$_4$. h$_3$ quantifies the skewness or asymmetry of the line, with positive values indicating a red-shifted tail (lower velocities) and negative a blue-shifted tail (higher velocities) of the Gaussian profile \citep{Marel1994}. h$_4$ quantifies the flatness/peakness (kurtosis) of the Gaussian line, with positive values associated with radial anisotropy and more negative values with tangential anisotropy \citep{Cappellari2004,Cappellari2017}. The h$_3$ maps show a clear anti-correlation with respect to the LOS velocity as expected for disk like type of rotation \citep[e.g][]{Krajnovic2008,Sande2017}. The h$_4$ maps look more patchy and noisier than the h$_3$ ones. M81, NGC~1052, NGC~4258, and NGC~5128 show a decrease in h$_4$ in the central $\lessapprox\, 0.3\arcsec$, while M104 and M94 show positive central values, suggesting more radial stellar orbits \citep[e.g][]{Cappellari2016}. 

\subsection{Gas kinematics fits}
\label{subsec:gas_kin}
The NIRSpec spectra allow us to simultaneously fit for several gas emission lines, including NIR coronal lines. Depending on their ionization potential, these lines serve as a probe for different physical processes in the vicinity of the central SMBH. From these lines we derive both kinematic maps and emission line flux maps; the latter are discussed in Section~\ref{subsec:flux_distribution}. 
To measure the gas kinematics,  we first subtracted the stellar spectrum extracted with {\sc pPXF} to obtain a ``gas only'' spectrum. This method has been used in previous works because it can provide better gas kinematic maps than fitting the stellar and gas kinematics simultaneously \citep[e.g][]{Neumayer2007}. We adopted a single-Gaussian fit to characterize the gas line profiles. Although this approach provides a robust first-order description of the velocity structure, it may oversimplify regions with complex gas kinematics. A more detailed analysis of the gas kinematics and dynamics will be presented in a follow-up work. 

We fit the six molecular hydrogen H$_2$ lines 1-0 S(3), 1-0 S(2), 1-0 S(1), 1-0 O(3), 1-0 O(5), 0-0 O(9) as a single kinematic component, with all lines fitted simultaneously. The individual kinematics of these lines are similar, and fitting them together produced cleaner velocity maps.  The rest of the emission lines are fitted independently, each as a single kinematic component.
We can identify NIR coronal lines such as [Mg IV] ($\lambda =4.48\,\mu {\rm m}$) in the aperture extraction spectra of galaxies such as M81, NGC~1052, NGC~4258, NGC~4395, M104, and NGC~5128, as illustrated in Figure~\ref{fig:example_spec_ReveaLLAGN}. Additionally, [Ar VI] ($\lambda = 4.5\,\mu {\rm m}$) is detectable in NGC~1052 and [Si VI] ($\lambda = 1.9\,\mu {\rm m}$) in NGC~4395 and NGC~5128. However, in the single-spaxel spectra of the innermost Voronoi bins, the S/N ratio is insufficient to detect these lines and determine their kinematics, except in the cases of NGC~4395 and NGC~5128. These two cases are discussed separately in Section~\ref{subsec:NGC4395},\ref{subsec:CentaurusA}), respectively, and the corresponding coronal line velocity maps are shown in Figure~\ref{fig:NGC5128_ions_maps}.

Figure~\ref{fig:gas_kinematic_maps} shows the LOS velocity maps for the molecular hydrogen lines as well as the two brightest recombination lines, Pa${\alpha}$ and Br${\gamma}$. We do not present the LOS velocity and velocity dispersion maps for the rest of the hydrogen recombination lines due to their low S/N, but we note that those with enough S/N show kinematics similar to Pa$\alpha$. We derive position angles and amplitudes for the gas LOS velocity maps in the same way as we did for the stellar kinematics above. We also show the direction of the radio jet in the gas kinematics and excitation maps, as its direction may coincide with features in the gas kinematics or emission that could indicate jet gas interactions. The radio jet or AGN driven outflows can produce shocks that preferentially heat the molecular gas relative to the recombination lines \citep[e.g.,][]{Ogle2007,Ogle2010,Guillard2012,Riffel2021}. This can help to visually identify regions where enhanced velocity dispersion may be due jet interactions. However, we caution that the possible link to the jet is based only on the alignment between the gas kinematic PA and the jet direction, and enhanced velocity dispersion can could trace other mechanisms, such as AGN-driven outflows.

\begin{figure*}
\includegraphics[width=.91\textwidth]{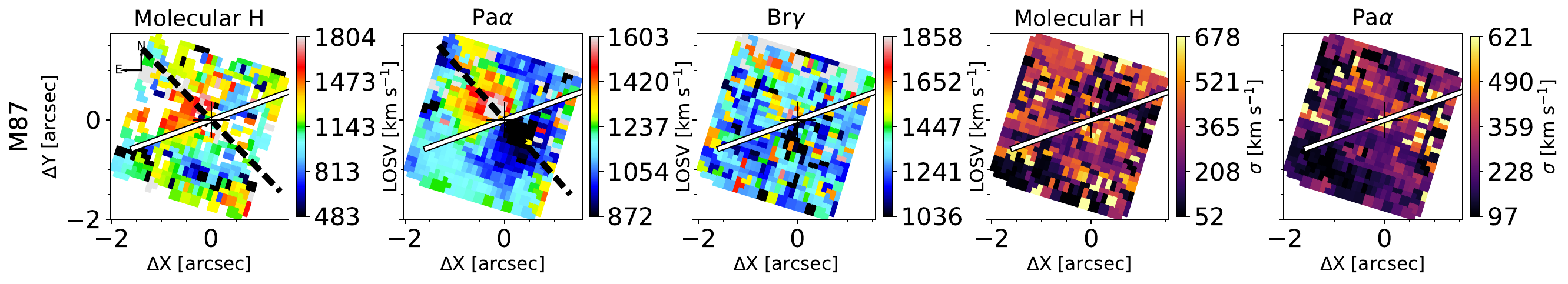}
\includegraphics[width=.91\textwidth]{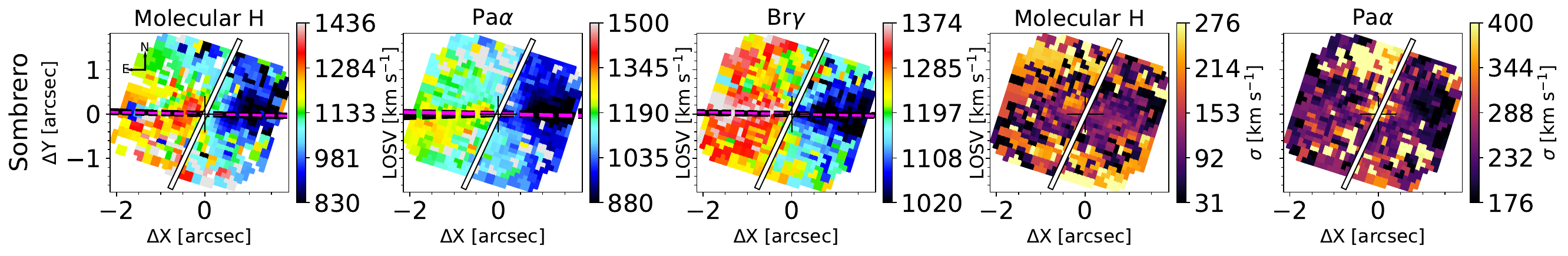}
\includegraphics[width=.91\textwidth]{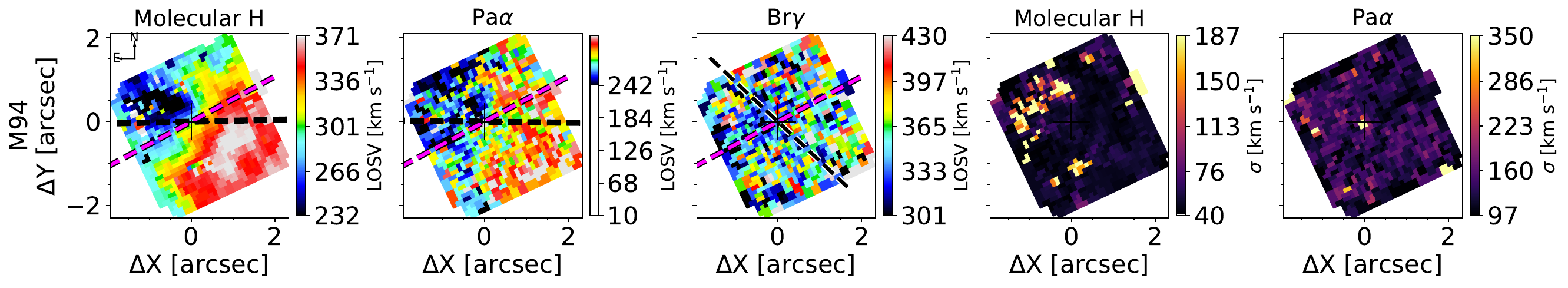}
\includegraphics[width=.91\textwidth]{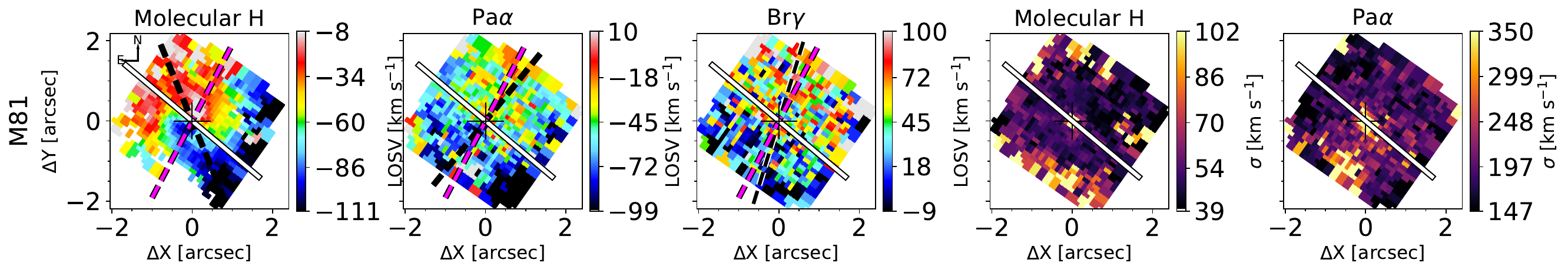}
\includegraphics[width=.91\textwidth]{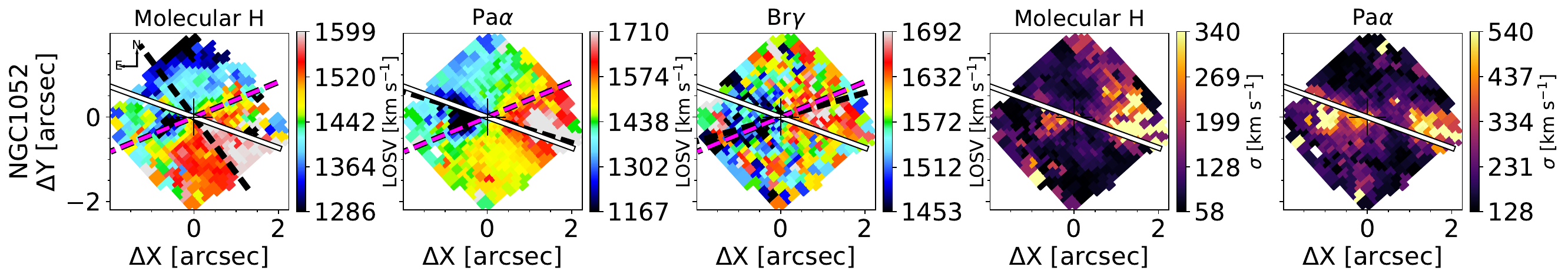}
\includegraphics[width=.91\textwidth]{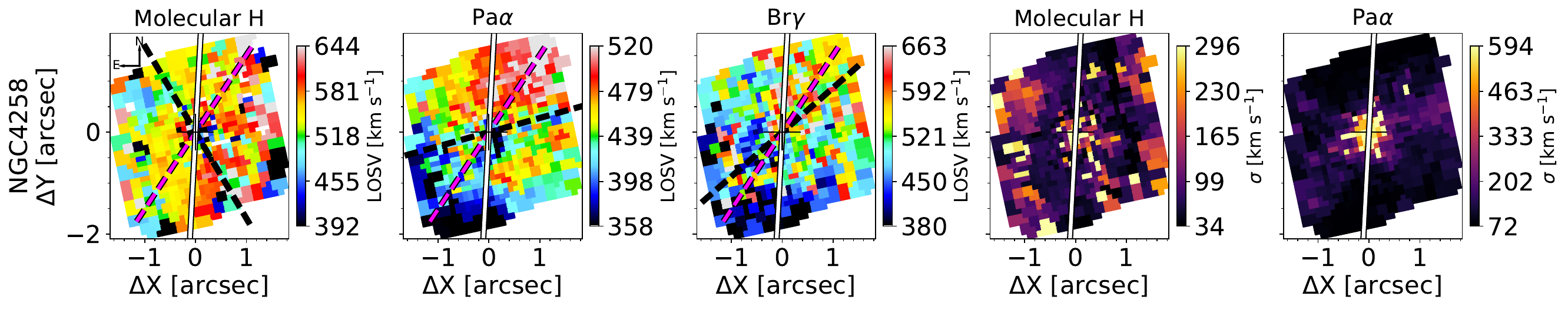}
\includegraphics[width=.91\textwidth]{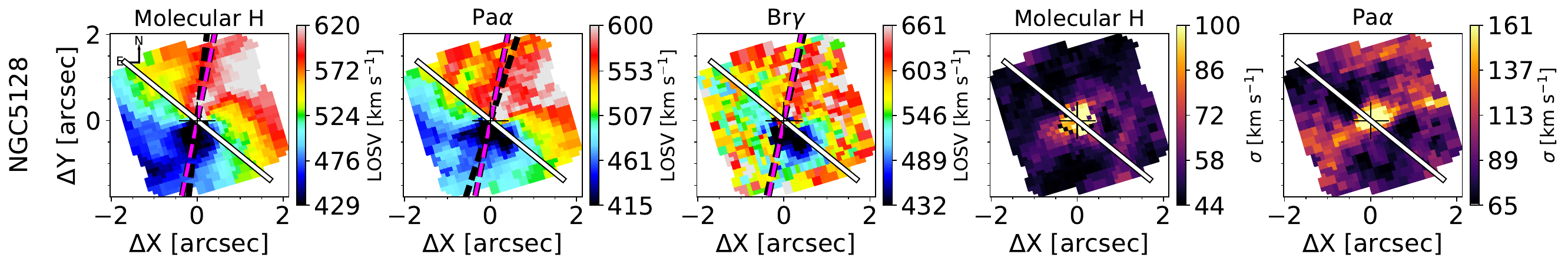}
\includegraphics[width=.91\textwidth]{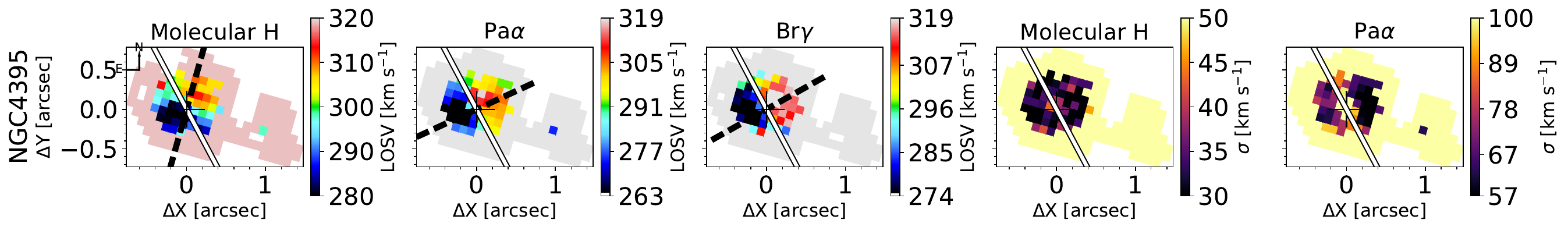}
\caption{kinematics maps of the molecular Hydrogen H$_2$ and Hydrogen recombination lines Pa$\alpha$ and Br$\gamma$. The first three panels show the LOS velocity and the last two the velocity dispersion. The dashed black line shows the best-fit gas kinematic PA. The solid white line marks the direction of the radio jet. \textbf{The dashed fuchsia line shows the stellar kinematic PA.} }
\label{fig:gas_kinematic_maps}
\end{figure*}

The H$_2$ lines show rotation similar to the stars, but with higher velocity amplitudes. This is consistent with the molecular gas being dynamically colder and more rotationally supported than the stellar component, which has a larger velocity dispersion and therefore more pressure supported. In galaxies like M81, M94, NGC~1052 and NGC~5128, the molecular gas motion seems to be decoupled from the stellar component, having a different kinematic PA. The decoupling suggests that the gas has not fully settled into the same dynamical plane as the stars, and it could instead originate from accretion, or be dominated by perturbations from AGN driven outflows and jet interactions.  In the specific case of NGC~5128 the gas probably was accreted during the last major merger event, as mentioned in previous studies \citet[e.g][]{Cappellari2009,Neumayer2010,Wang2020,Dumont2025}.

\subsection{Flux and line ratios maps}
\label{subsec:flux_distribution}
In addition to the kinematics we also present the distribution of line intensities and ratios in the nuclear regions. Figure~\ref{fig:flux_maps} shows the flux maps for the molecular hydrogen gas (i.e the sum of the five molecular hydrogen lines) as well as the Pa$\alpha$ and Br$\gamma$ lines that trace the ionized gas. For the case of NGC~4395 and NGC~5128 we also provide flux maps in Figure~\ref{fig:NGC5128_ions_maps} for the coronal lines for which we could obtain kinematic information. The two right panels in Figure~\ref{fig:flux_maps} show the H$_2$/Br$\gamma$ line ratio map and a colored version of the same map distinguishing the different excitation mechanisms in four different colors. Following \citet{Riffel2021}, we classify the excitation as star-formation for H$_2$/Br$\gamma<0.4$ (blue), low AGN photoionization for $0.4 \leq$ H$_2$/Br$\gamma$ $\leq2.0$ (orange), high-AGN photoionization for $2.0<$ H$_2$/Br$\gamma\leq6.0$ (red), and shock excitation for H$_2$/Br$\gamma>6$ (green).

The H$_2$/Br$\gamma$ line ratio maps are constructed from the prominent ro-vibrational hydrogen H$_2$ 1-0 S(1) line flux at $2.12\,\mu {\rm m}$ which is close to the  Br$\gamma$ ionized hydrogen line at 2.1661 $\mu {\rm m}$. Their proximity in wavelength makes this comparison relatively insensitive to differential dust extinction. The H$_2$/Br$\gamma$ ratio has been used previously as a diagnostic tool to distinguish between different excitation mechanisms in LLAGN \citep[e.g][]{Reunanen2002,Rodriguez2004,Colina2015,Riffel2013,Riffel2021}. This line ratio is effective because Br$\gamma$ is primarily excited by UV photons from young stars in star forming regions or AGN continuum, while the ro-vibrational line H$_2$ can also be excited by shocks \citep{Hollenbach1989} and X-ray photons \citep{Maloney1996}. However, it should be noted that the H$_2$/Br$\gamma$ line ratio thresholds are empirically derived and that they are not as well understood as the optical line ratios \citep{Rodriguez2004}. This diagnostic also cannot distinguish whether the shocks are caused by interactions between the jet and the gas or by AGN-driven outflows.

We find a mean H$_2$/Br$\gamma$ value of 1.6 which is consistent with the range for AGN excitation, with a minimum value of H$_2$/Br$\gamma \sim 0.8$ in M81 and a maximum of H$_2$/Br$\gamma \sim 3.7$ in NGC~5128. We also observe higher H$_2$/Br$\gamma$ values at larger radii compared to the center in three systems, NGC~5128, NGC~4395 and NGC~4258. This could be interpreted as evidence for Br$\gamma$ being more centrally concentrated and excited by the LLAGN, while the H$_2$ is more extended and probably excited by other processes such as stellar winds and/or shocks. 

\begin{figure*}
\includegraphics[width=.9\textwidth]{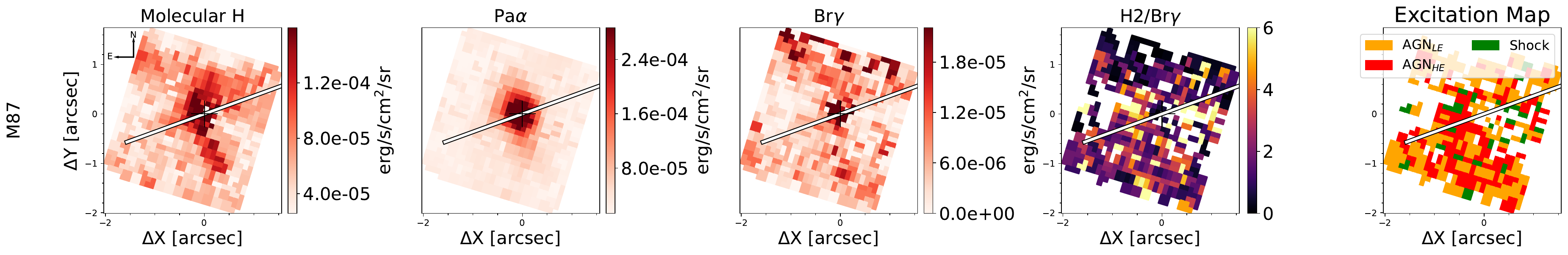}
\includegraphics[width=.9\textwidth]{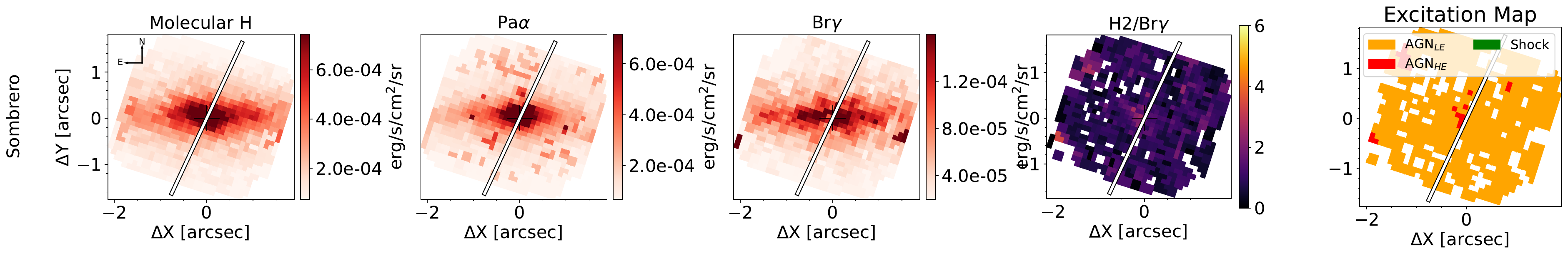}
\includegraphics[width=.9\textwidth]{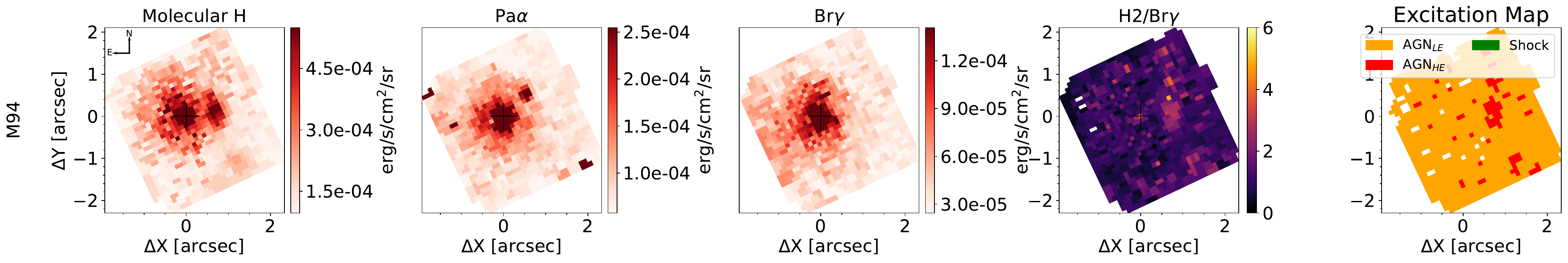}
\includegraphics[width=.9\textwidth]{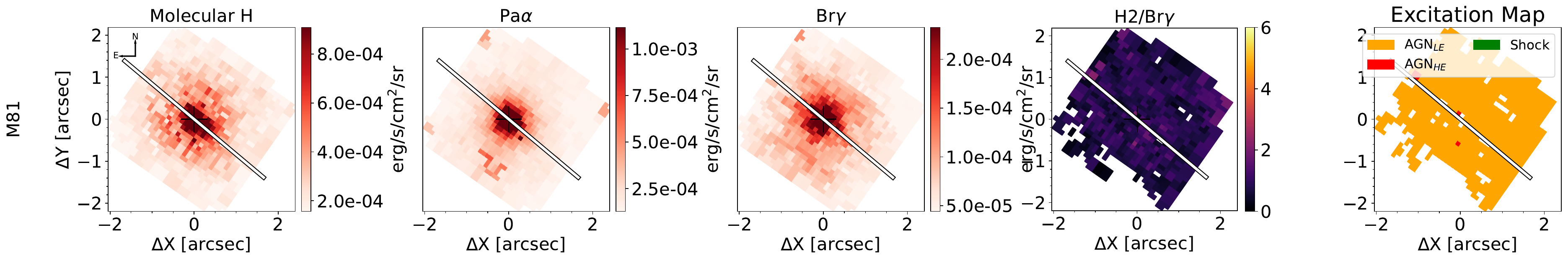}
\includegraphics[width=.9\textwidth]{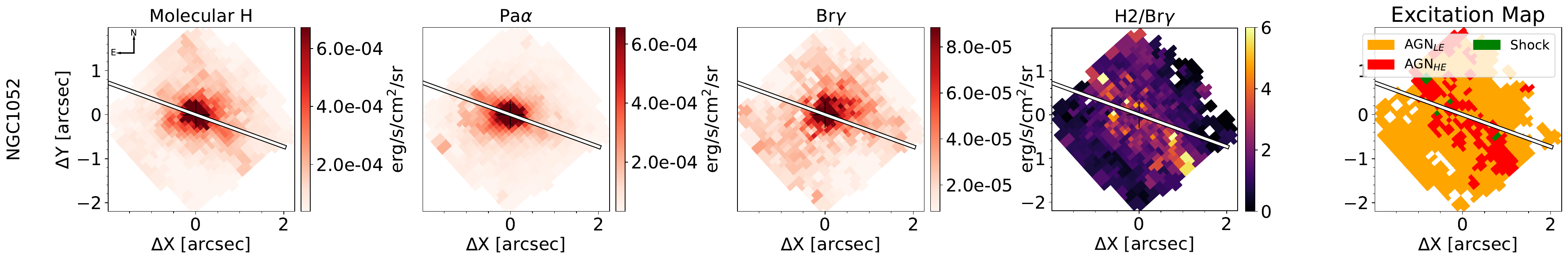}
\includegraphics[width=.9\textwidth]{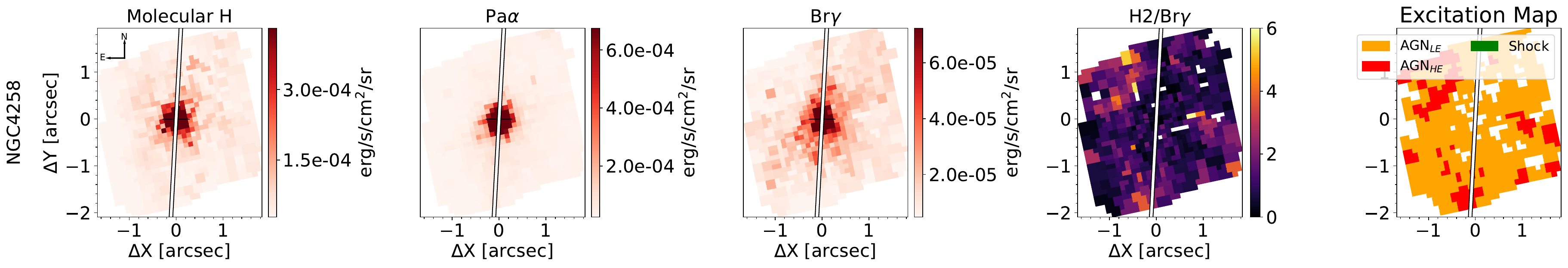}
\includegraphics[width=.9\textwidth]{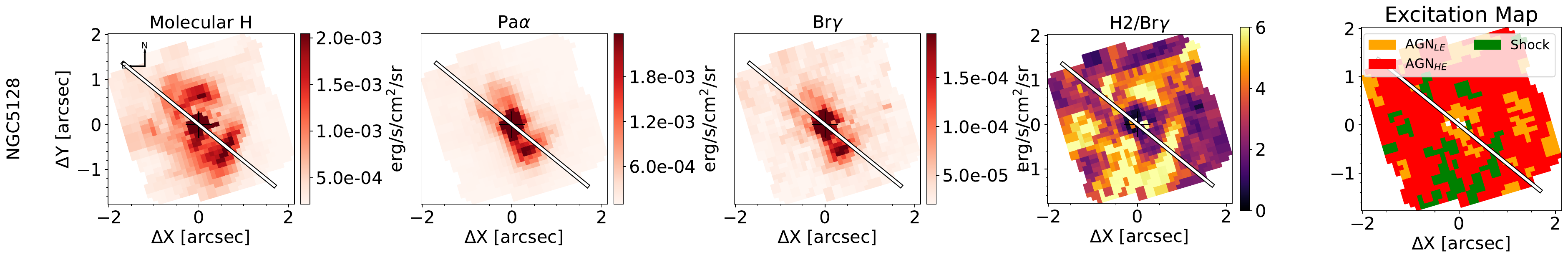}
\includegraphics[width=.9\textwidth]{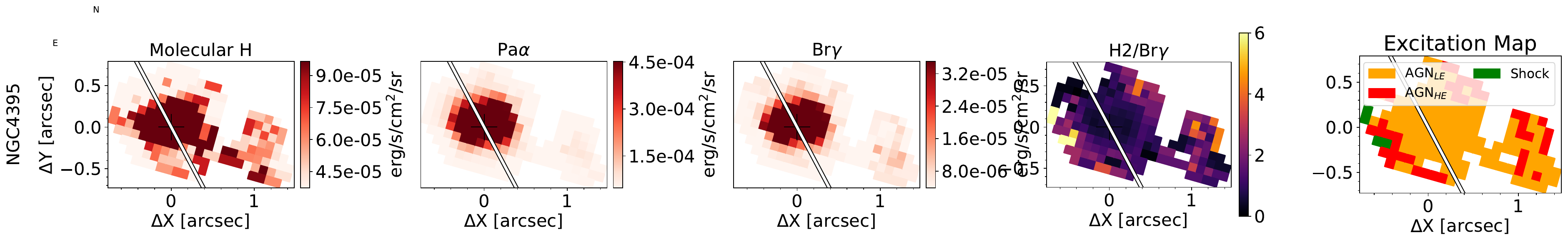}
\caption{Flux and line ratio maps. The left three panels show the flux maps for the H$_2$ 1-0 S(1),$Pa\alpha$, and Br$\gamma$ lines. The right two panels show the H$_2$/Br$\gamma$ line ratio map and a colorized version highlighting the different excitation mechanism. Values of H$_2$/Br$\gamma \leq 0.4$ (blue) indicate star forming regions,  $0.4<$ H$_2$/Br$\gamma \leq 2.0$ low-AGN excitation (orange), $2.0<$ H$_2$/Br$\gamma \leq 6.0$ high-AGN and H$_2$/Br$\gamma$ above 6.0 shock excitation (green). The white line shows the direction of the radio jet.}
\label{fig:flux_maps}
\end{figure*}

In what follows, we provide a more detailed discussion of the stellar and gas kinematics as well as the line intensity and line ratio maps for each galaxy. 

\subsection{M87}
\label{subsec:M87}
M87 (NGC~4486) hosts one of the most massive black holes known, M$_{\rm BH} \sim 6.5 \times 10^9\,M_\odot$ \citep{Gebhardt2011,Liepold2023} and has been famously observed as part of the Event Horizon Telescope sample \citep{EHT2019}. The M87 black hole is the origin of a relativistic jet oriented at PA$_{jet} \sim -70^\circ$ east of North \citep[e.g.][]{Walker2018}. 

The \textbf{stellar kinematics} of M87 are known to have small LOS velocity amplitudes ($\Delta_{\rm LOSV}$) of $\sim 45$ km s$^{-1}$ and a complex structure \citep{Emsellem2014,AlAmri2025}. In our M87 data, stellar features, such as the CO bandhead, are not visible in the nuclear regions (radii $\lesssim0.4\arcsec$) because of M87's low surface brightness. This limitation is further exacerbated by the presence of wiggles in the nuclear region, preventing us from deriving stellar kinematics. The same issue prevented \citet{AlAmri2025} from deriving stellar kinematics within the inner $0.5\arcsec$ of M87, despite using deeper NIRSpec data than in this work. 

The \textbf{ionized and molecular gas kinematics} display disk rotation (Figure~\ref{fig:gas_kinematic_maps}, top-row). The Pa$\alpha$ velocity map shows a well defined gradient along a PA of 42$^\circ$ east of north, while the molecular hydrogen velocity gradient is oriented along a PA 54$^\circ$ east of north. The PA for Pa$\alpha$ is in good agreement with the PA of $45^{\circ}$ derived for H$\alpha$ by \citet{Walsh2013}. The H$_2$ LOS velocity map has a more irregular morphology compared to the ionized gas, with hints of turbulence or non-circular motions. The molecular gas velocity dispersion reaches up to $\sim700~\mathrm{km~s^{-1}}$ near the jet axis and also along the red shifted side of the map, which could suggest a strong dynamical perturbation. In contrast, the Br$\gamma$ line is weaker, and its velocity map does not show a coherent rotation pattern; therefore, we do not derive a kinematic PA for this line.
The \textbf{flux maps} (Figure~\ref{fig:flux_maps}, top-row) for the ionized gas look compact and peak at the center, while the flux map for molecular hydrogen is more extended. The H$_2$ emission has a linear structure that is almost perpendicular to the direction of the radio jet and does not seem to be correlated with any visible structure in the ionized gas. 
The H$_2$/Br$\gamma$ line ratio is highly impacted by the low signal-to-noise of the H$_2$ 1-0 S(1) line, but it reveals that photo-excitation from the central AGN is the primary source of excitation over mechanical heating from the jet in the inner $\sim100$~pc. The line ratio map shows small sections of mechanical shocks that excite the gas away from the direction of the jet. 

\subsection{M104}
\label{subsec:Sombrero}

The M104 galaxy, also known as Sombrero or NGC 4594, is a nearby elliptical galaxy (D=9.8 Mpc) that has been thoroughly studied at many wavelengths due to its proximity. M104 hosts a central supermassive black hole \citep[e.g][]{Kormendy1996A} and has a prominent dust lane and a large scale stellar disk \citep[e.g][]{Kormendy1988}. M104 has a parsec scale compact radio emission that is associated with the AGN \citep{Hada2013}. 

The \textbf{stellar kinematics} of M104 have been previously studied using ground-based observations from Gemini NIFS IFU, which revealed a disk like rotation with a LOS velocity amplitude of $\Delta_{\rm LOSV}$ of $\sim 240\pm7$ km s$^{-1}$ \citep{Menezes2015}. Our JWST/NIRSpec data show consistent stellar disk kinematics with a disk rotation (Figure~\ref{fig:stellar_kinematic_maps}, top-row) with a $\Delta_{\rm LOSV}$ of $\sim 238$ km s$^{-1}$ and a PA of $88.8^\circ$ east of north. The stellar velocity dispersion shows an increase towards the center of the disk with a peak of $\sigma=267\pm9$ km s$^{-1}$. The $\sigma$ shows a peanut shape along the disk, and not a single circular central peak. \citet{Menezes2015} found an off-center $\sigma$ peak, shifted $0.21\arcsec$ east of the center, with a value of a $\sigma$ peak of $\sim 272\pm9$ km s$^{-1}$. The apparent difference between these results may be due to the higher sensitivity of the JWST/NIRSpec data, which allow us to resolve the extended peanut-shaped structure that may have appeared as a single off-centered peak in the Gemini/NIFS observations.

The stellar velocity dispersion also increases away from the stellar disk (north-south). This suggests that the structure of the nuclear region of M104 consists of a kinematically cold stellar disk rotating west to east embedded in a kinematically hot bulge \citep{Menezes2015}. The h$_3$ map is anti-correlated with the LOS velocity, as expected for a disk rotation. The h$_3$ switches signs in the inner $\sim 0.15\arcsec$ along the disk. The maps of the higher-order Gauss–Hermite moments are more patchy than those of the LOS velocity or the velocity dispersion, and the apparent sign reversal in h$_3$ may simply be due to noise. However, it might also indicate the presence of an additional kinematic component in the innermost region of M104, such as a nuclear star cluster. The h$_4$ map is the noisiest among all the maps and exhibits an increase toward more positive values at the center, which is associated with the central black hole \citep{Cappellari2017}. 

The \textbf{ionized and molecular gas kinematics} (Figure~\ref{fig:gas_kinematic_maps}, second-row) follow the stellar rotation, showing a similar disk rotation with kinematic PA of 87$^{\circ}$, 95$^{\circ}$ and 89$^{\circ}$ for the H$_2$ LOS , Pa$\alpha$ and Br$\gamma$. The overall H$_2$ LOS velocity maps is similar to the ionized gas but show bad fits in the outer regions. 
The LOS velocity maps of the ionized gas traced by Pa$\alpha$ and Br$\gamma$ are closely aligned with the stellar kinematics. 
The molecular gas velocity dispersion shows a profile similar to that of the stars, characterized by a central peak and decreasing values at larger radii along the disk. The ionized gas velocity dispersion also shows enhanced values away from the disk plane like the stars. However, these values are likely affected by unreliable spectral fits and should be interpreted with caution.

The \textbf{flux maps} for the molecular and ionized gas (Figure~\ref{fig:flux_maps}, second-row) show a extended emission from the center all the way to the outer parts of the disk. The line ratio map reveals AGN-driven photoionization throughout the field of view, with evidence for a stronger ionizing radiation field at the very center.

\subsection{M94}
\label{subsec:M94}
M94 or NGC~4736 is a close face-on spiral galaxy and one of the nearest LINERS at a distance of 5 Mpc \citep{Constantin2012}. The presence of a radio jet in M94 is debated, where some studies require a jet component to fit its broad SED, but no radio emission has been detected \citep{Oers2017}.

The \textbf{stellar kinematics} of M94 (Figure~\ref{fig:stellar_kinematic_maps}, second-row) reveal a slow rotation with only a LOS velocity amplitude of $\Delta_{\rm LOSV}$ of $\sim 20\pm7$ km s$^{-1}$. The rotation pattern is consistent with a disk like rotation along a kinematic PA of $117^\circ$ east of north. In contrast to the typical central rise expected in the presence of a supermassive black hole, the stellar velocity dispersion map shows a decrease toward the nucleus. The stellar dispersion drops from an average $\sigma\approx98$ km s$^{-1}$ to $\sigma\approx80$ km s$^{-1}$ in the inner $\sim0.35\arcsec$. However, a modest central peak is detected within the innermost $\sim0.05\arcsec$, where the dispersion increases to $\sigma\approx92$ km s$^{-1}$, consistent with the influence of the central black hole. 

The h$_3$ map is patchy, but shows an anti-correlation with the LOS velocity, as expected for a rotating disk. The h$_4$ map indicates predominantly radially anisotropic stellar orbits across the field of view and with and increase to higher values (h$_4\sim0.3$) at the center. Interestingly, the peak in $h_4$ is offset from the photometric center by $\sim0.1\arcsec$. This offset spatially coincides with structures observed in the molecular and ionized gas flux maps. The H$_2$ emission displays a double-peaked morphology, with one peak located east and another west of the nucleus. The eastern peak aligns with the location of the enhanced $h_4$ values. The double peak emission is also observed in the ionized gas maps but is shifted by about $45^\circ$. The line ratio map discards this region as being a star forming region, having line ratios of about $\sim4$ well consistent with high AGN photoionization. 

The \textbf{ionized and molecular gas kinematics} (Figure~\ref{fig:gas_kinematic_maps},third-row) show an overall rotation from east to west twisted by $\sim 20^{\circ}$ with respect to the stellar rotation. The molecular gas presents several twists in the kinematic PA in the north-south direction. The gas kinematics is decoupled from the stellar kinematics. The LOS velocity for the molecular gas has a $\Delta_{\rm LOSV}$ of $\sim 60$ km s$^{-1}$ and a PA of $\sim91^\circ$ east of north. The observed twisting in the molecular gas rotation may arise from a precessing disk, in which orbital orientations vary with radius \citep{Wong2000}, or alternatively from a tilted or warped disk configuration.
The ionized gas shows a rotational pattern similar to that of the molecular gas with a kinematic PA of $\sim89^\circ$ east of north. The velocity dispersion map of Pa$\alpha$ shows a clear increase toward the nuclear region.

Taken together, the gas kinematics in M94 are complex and cannot be adequately described by a single, simple rotating disk, indicating the presence of multiple kinematic components or a strongly non-axisymmetric gas configuration in the nuclear region.

\subsection{M81}
\label{subsec:M81}
M81 hosts one of the closest low-luminosity AGN at a distance of 3.61 Mpc (at a distance similar to NGC~5128), and is the closest target in the ReveaLLAGN sample. M81 exhibits a radio jet with a PA$_{jet}$ $\sim50^\circ$ \citep{Markoff2008}. The radio jet and the LLAGN are powered by the central supermassive black hole of $6.5\times10^{7}$ M$_{\odot}$ \citep{Kormendy2013}. 

The \textbf{stellar kinematics} map of M81 (Figure~\ref{fig:stellar_kinematic_maps}, third-row) shows a slow rotation with a kinematic PA=$152^\circ$ east of north, consistent with previous works \citep{SchnorrMuller2011}, and aligned with the galaxy's photometric major axis \citep{Goad1976}. The stellar rotation for M81 is consistent with a disk type rotation, with LOS velocity amplitude of $\Delta_{\rm LOSV}$ of $\sim 35$ km s$^{-1}$. The stellar LOS velocity from our JWST/NIRSpec is in good agreement with the ground based observations for M81 from Gemini/GMOS \citep{SchnorrMuller2011}. Based on the observed morphology of M81 and assuming trailing spiral arms, \citet{SchnorrMuller2011} conclude that the south-west side of the galaxy is the near side, while the north-east side is the far side. 

The stellar velocity dispersion $\sigma$ increases towards the center with a peak of $222.1\,\pm\,10.4$ km s$^{-1}$ within the central $0.1\arcsec$. The higher order Gauss-Hermite moments show modest amplitudes, with h$_3$ clearly anti-correlated with the LOS velocity as expected for disk-like rotation. The h$_4$ has a small peak in the inner $\leq0.05\arcsec$ indicative of the presence of a black hole. This peak is surrounded by a ring of lower h$_4$ values, which may indicate the presence of an additional kinematic component or a more complex orbital structure.

The \textbf{molecular and ionized gas kinematics} maps (Figure~\ref{fig:gas_kinematic_maps}, fourth-row) reveal rotation with a predominantly north-south velocity gradient and decoupled from the stellar rotation. The molecular gas has an overall kinematic PA of $23^\circ$ east of north, corresponding to a $\sim130^\circ$ misalignment with respect to the stellar disk. The molecular gas also presents twist in its kinematic PA. 
The ionized gas LOS velocity map is patchy and seems to align with that of the stellar rotation having also kinematic PA of $152^{\circ}$. The misalignment between the molecular gas and the stars could be indicative of gas accretion into the nuclear regions.

The velocity dispersion of the molecular gas is generally low across the field-of-view, with a central peak of $\sigma\approx90$ km s$^{-1}$. Elevated dispersion in the molecular gas is observed along the direction of the radio jet in the near-side of the galaxy (south-west), suggesting a local interaction between the jet and the molecular gas. The ionized gas displays a similar spatial trend but with significantly higher dispersion values, reaching $\sim400$ km s$^{-1}$ within the inner $0.1\arcsec$, consistent with more turbulent or disturbed ionized material.

The \textbf{flux maps} for both the molecular and ionized gas (Figure~\ref{fig:flux_maps},fourth-row) are centrally peaked with the molecular gas having a more extended emission. Line ratio maps of H$_2$/Br$\gamma$ for M81 suggest AGN photoionization as the primary excitation mechanism in the nuclear region, with no evidence of shocks playing a dominant role.

\subsection{NGC~1052}
\label{subsec:NGC1052}
NGC~1052 is a nearby elliptical galaxy hosting what could be described as a prototypical LINER \citep{Ho2008}. Despite being an elliptical galaxy, it has a dense nuclear accretion disk, making it one of the only two elliptical galaxies containing a very luminous maser \citep[e.g][]{Braatz1996}. NGC~1052 is also known for hosting a two sided radio jet at a PA$_{jet}\approx70^\circ$ \citep{Kadler2004b}. 

The \textbf{stellar kinematics} for NGC~1052 (Figure~\ref{fig:stellar_kinematic_maps}, fourth-row) show a clear disk rotation with a kinematic PA $=112^\circ$ east of north. The LOS velocity of NGC~1052 has a mean velocity of $1448.7$ km s$^{-1}$ and a rotation amplitude of $\Delta_{\rm LOSV}$ of $\sim 90$ km s$^{-1}$. The stellar velocity dispersion is centrally peaked with an average value of $\sim 300$ km s$^{-1}$ in the inner $0.1\arcsec$. The h$_3$ is anti-correlated with the LOS velocity supporting the idea of a disk rotation, with values ranging from $-0.05$ to $0.03$. The h$_4$ map shows a small east to west gradient from positive to negative values, or from radial to tangential anisotropy. However, the overall values of h$_4$ are relatively small with amplitudes |$h_4|\lesssim0.02$

The \textbf{molecular and ionized gas kinematics} (Figure~\ref{fig:gas_kinematic_maps}, fifth-row) both show signs of disk rotation with a primordially east-west velocity gradient. The LOS velocity map for the molecular hydrogen shows a disk like rotation with a kinematic PA of $57^\circ$ east of north, but with signs of non-rotation along the radio yet in the south-west side of the galaxy. In this region, higher LOS velocity amplitudes coincide spatially with larger molecular gas velocity dispersion, suggesting a dynamical interaction between the jet and the molecular gas. The ionized gas velocity field displays a relatively regular rotation pattern, with a kinematic position angle of $79^\circ$ derived from the Pa$\alpha$ emission and no clear large-scale kinematic disturbances aligned with the radio jet. However, the Pa$\alpha$ velocity dispersion map shows elevated values, both in the central region and along the projected jet direction, reaching $\sigma \approx 450$ km s$^{-1}$. These high dispersions may indicate the presence of multiple kinematic components, such as a combination of rotation and jet-driven outflows, which are not fully captured by a single-Gaussian fit. Similar complex ionized gas kinematics, requiring multiple components to reproduce the emission-line profiles, have been reported in HST/STIS observations of H$\alpha$ and [N\,II] \citep{Walsh2008}.

The \textbf{flux and line ratio maps} (Figure~\ref{fig:flux_maps}, fifth-row) confirm the idea of jet gas interaction. The flux maps show centrally concentrated flux, but with tail-like extensions roughly in the direction of the jet. The H$_2$/Br$\gamma$ line ratio map indicates that AGN photoionization dominates the excitation of the nuclear gas. However, localized regions along the jet direction (at PA $\sim70^\circ$ east of north) show line ratios consistent with an additional contribution from mechanical shock excitation.

\subsection{NGC~4258}
\label{subsec:NGC4258}

NGC~4258 is a nearby (7.2 Mpc) well study Seyfert galaxy hosting a powerful water maser with a radio jet extending from sup-parsec scale to kilo-parsec scales \citep{Cecil2000} with a PA$_{jet}\approx-3$ east of north. The central water maser disk has been used to measure the mass of the supermassive black hole, first by \citet{Miyoshi1995} and more recently with improved precision by \citet{Reid2019}, giving $M_{\rm BH} = 3.98 \times 10^{7}$ M$_\odot$. 

The \textbf{stellar kinematics} (Figure~\ref{fig:stellar_kinematic_maps}, fifth-row) derived from our JWST/NIRSpec observations reveal one of the clearest cases of disk like rotation in our sample, with a stellar kinematic PA $=146^\circ$ east of north. The LOS velocity has an amplitude $\Delta_{\rm LOSV}$ of $\sim 60$ km s$^{-1}$. The stellar velocity dispersion shows a rise towards the center, from an average of $\sigma\approx 80$ km s$^{-1}$ in the outer $0.5\arcsec$ to $\sigma\approx 205$ km s$^{-1}$ within the inner $0.1\arcsec$. The h$_3$ map shows values between $-0.2$ and 0.2 and displays a clear anti-correlation with the LOS velocity, consistent with expectations for disk rotation. The h$_4$ values remain close to zero across the field of view, indicating that the line profiles are predominantly Gaussian and the stellar orbits largely isotropic, except within the innermost $\sim 0.1\arcsec$, where an increase toward higher h$_4$ values is observed, attributable to the influence of the central black hole.

\textbf{Molecular and ionized gas kinematics} maps (Figure~\ref{fig:gas_kinematic_maps}, sixth-row) show a disk like rotation aligned with the stellar rotation, but with localized deviations. The molecular gas velocity map reveal non-circular motions along the southern side of the radio jet. The radio jet is known to precess, and in the past the southern jet was oriented more toward the west than it is at present \citep{Cecil2000}. This precession may explain why the enhancement in LOS velocity and velocity dispersion of the molecular gas is located slightly west of the current jet axis. The derived kinematic PA for the molecular gas is $24^{\circ}$. The LOS velocity map for the Pa$\alpha$ seems to be aligned with the stellar rotation, showing a similar kinematics PA of $152^\circ$ east of north. The ionized gas rotation also appears to be influenced by the radio jet in the same sense as the molecular gas, but to a lesser extent. Its velocity dispersion does not exhibit an elongation along the jet and is instead centrally concentrated.

The \textbf{flux maps} (Figure~\ref{fig:flux_maps}, sixth-row) show a centrally concentrated flux distribution for the molecular and ionized gas. However, there are small tails of extended molecular gas emission in the south-west direction that could be linked with the radio jet. Regardless, the line ratio map shows the central AGN emission as the main excitation mechanism in the nuclear region, surprisingly with no shocks along the jet direction. 

\subsection{NGC~5128}
\label{subsec:CentaurusA}
NGC~5128 (Centaurus~A) is the nearest AGN hosting a radio loud jet (3.6 Mpc). It host a supermassive black hole at its center with a mass of $4.5 -5.5\times10^7$ M$_\odot$ \citep{Neumayer2007,Cappellari2009}. The radio loud jet is oriented $51^\circ$ east of north \citep{Jones1996}.

The \textbf{stellar kinematics} map of NGC~5128 (Figure~\ref{fig:stellar_kinematic_maps}, sixth-row) reveals a slowly rotating, disk like stellar kinematics with a kinematic PA $=178^\circ$ east of north. The LOS velocity has an amplitude of $\Delta_{\rm LOSV}\approx 30$ km s$^{-1}$. This $\Delta_{\rm LOSV}$ agrees with the rotational amplitude derived from ground-based data in \citet{Cappellari2009}. The stellar velocity dispersion increases toward the center as expected for a system hosting a central black hole, and peaks at $\sigma\sim 246$ km s$^{-1}$. The h$_3$ moment is anti-correlated with the LOS velocity, while h$_4$ shows no prominent trend and remains close to zero ($-0.02\leq${\rm h$_4$}$\leq0.01$).

The \textbf{ionized and molecular gas kinematics} maps (Figure~\ref{fig:gas_kinematic_maps}, seventh-row) show a counter rotating gas with respect to the stellar disk. The molecular gas kinematics show no signs of being altered by the radio jet and with a kinematic PA of $171^{\circ}$. The Pa$\alpha$ and Br$\gamma$ velocity maps also reveal a disk like rotation with a kinematic PA$= 162^\circ$ and $169^\circ$ east of north, respectively. The ionized gas species also display non rotational motion around the south lobe of the radio jet as mentioned in previous studies \citep[e.g][]{Neumayer2007,Dumont2025}. The velocity dispersion map for the gas shows ring like structure with high values of dispersion and a central peak. Comparing our results to the gas kinematics from \citet{Neumayer2007}, we can associate the inner peak in the gas LOS velocity to the gravitational influence of the central supermassive black hole.

Out of the full sample, NGC~5128 and NGC~4395 are the only galaxies where we can obtain kinematics for coronal lines. In the case of NGC~5128 for [Si VI] at $1.96\,\mu {\rm m}$, [Mg VIII] at $3.02\,\mu {\rm m}$ and [Ar VI] at $4.52\,\mu {\rm m}$ (Figure~\ref{fig:NGC5128_ions_maps}). The velocity maps for the coronal lines are all similar and show a velocity gradient with a PA of $\sim 97^\circ$ east of north, which is almost 90$^\circ$ to the rotation of the molecular hydrogen. The velocity gradient for the coronal lines aligns well with the radio jet, as is the case for the non-rotational signatures seen in the Pa$\alpha$ and Br$\gamma$. Their flux maps also show extended emission and a similar structure to that of the ionized gas, with higher emission of gas along the jet. This points to an excitation driven by the mechanical ionization of the radio jet. 
 
The \textbf{flux maps} (Figure~\ref{fig:flux_maps}, seventh-row) for the molecular hydrogen show a compact source at the center surrounded by large ring-like structures of gas. The ionized gas and coronal lines show similar extended emission, but are concentrated around the direction of the jet. The H$_2$/Br$\gamma$ line ratio map shows the AGN photoionization as the primary source of excitation throughout the field-of-view. However, large regions of shock excitation are seen along the south part of the radio jet, and also corresponding with the brightest parts of the ring-like clouds. 

The coronal emission is potentially elongated along the direction of the radio jet as seen in Figure~\ref{fig:NGC5128_ions_maps},  while our H$_2$/Br excitation map shows regions with line ratios consistent with shock excitation along the same direction. This spatial connection  suggests that interactions between the radio jet and the surrounding gas may contribute to the excitation of the extended coronal emission as observed in other LLAGN \citep{May2018, Rodriguez2025}.

\begin{figure*}
\includegraphics[width=.5\textwidth]{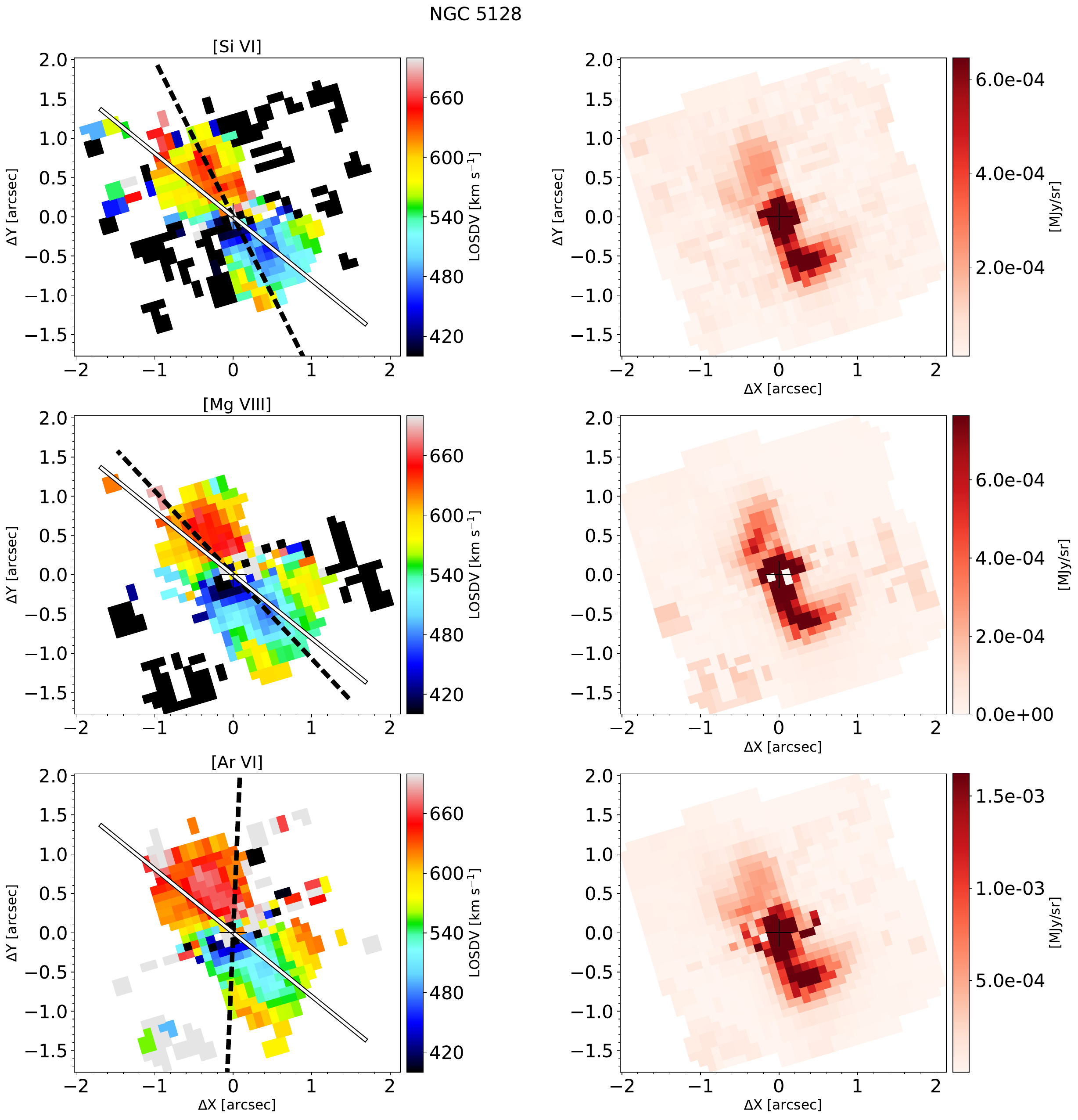}
\includegraphics[width=.5\textwidth]{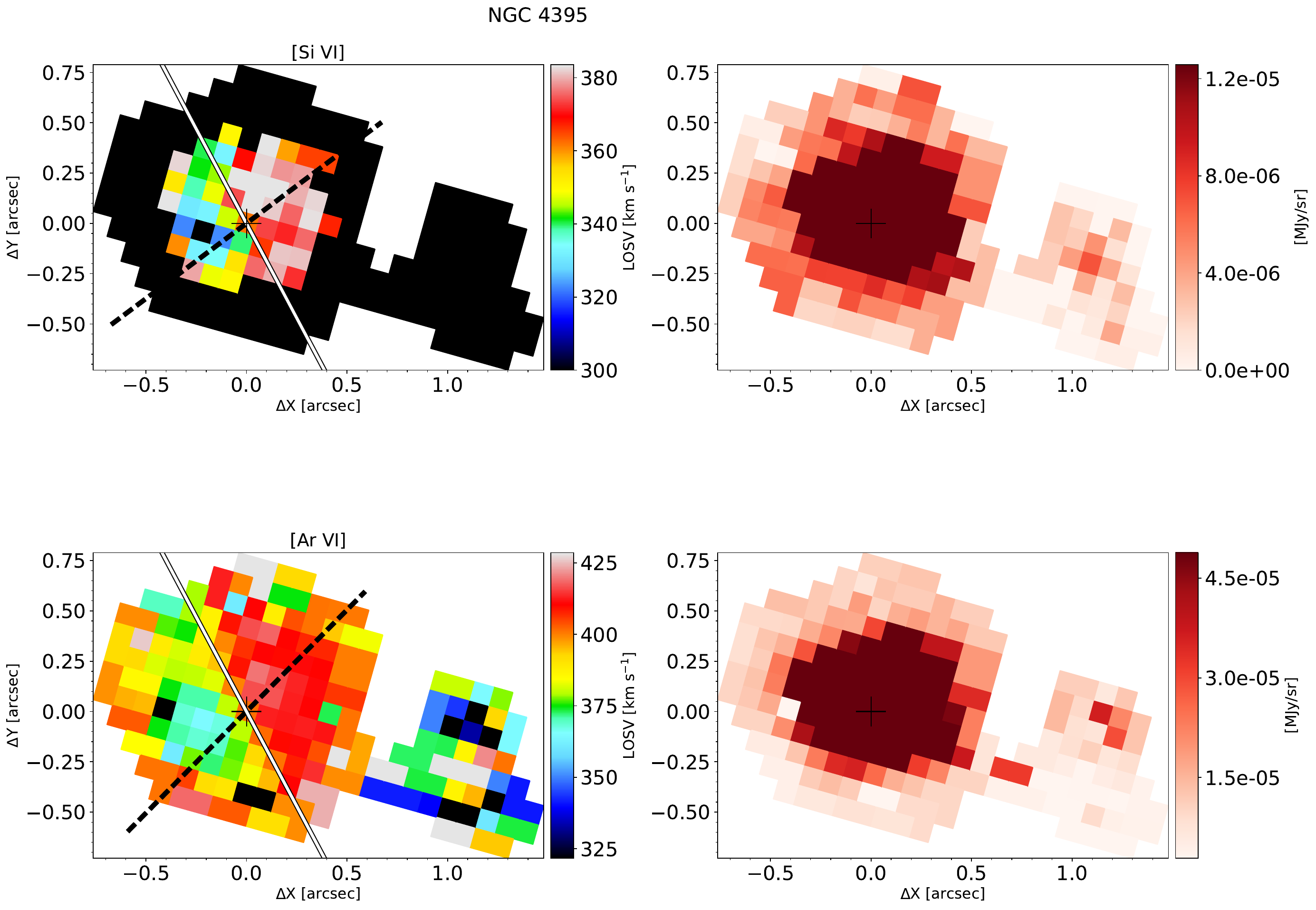}
\caption{Velocity and flux maps for the coronal lines observed in NGC~5128 (left) and NGC~4395 (right).}
\label{fig:NGC5128_ions_maps}
\end{figure*}

\subsection{NGC~4395}
\label{subsec:NGC4395}
NGC~4395 is a nearby bulgeless early type galaxy hosting one of the lowest mass supermassive black holes $\sim10^{4-5}$ M$_{\odot}$ \citep{Filippenko2003}. Recent radio observations taken with the VLA presented in \citet{Nandi2023} show a two-sided asymmetric radio blobs of jet-ISM interactions in the inner 10 pc with a PA$_{jet}\approx72^\circ$ east of north.

We could not fit the \textbf{stellar kinematics} for NGC~4395. Given the bulgeless nature of NGC~4395 and the very faint stellar continuum there is no visible CO bandhead in the JWST/NIRSpec data, nor other stellar features that we could use to derive stellar kinematics. We also attempted to extract a single stellar kinematic measurement from spectra using different apertures, including a nuclear aperture and an annular aperture subtraction, but no stellar features were recovered and we were therefore unable to derive reliable stellar kinematics. This limitation in studying NGC~4395's stellar kinematics has also been encountered in previous works using NIR IFU data from Gemini/NIFS \citep{Denbrok15,Brum2019}. To date, the only stellar kinematic measurement available for NGC~4395 comes from a slit spectrum of the Ca II triplet \citep{Filippenko2003}, where the AGN contribution could be less dominant than at NIR wavelengths or could be due to AGN variability \citep{Minezaki2006A}.

The \textbf{ionized and molecular gas kinematic} maps (Figure~\ref{fig:gas_kinematic_maps}, bottom-row) are limited to the inner $\sim0.5\arcsec$ were we can achieve good S/N to get gas kinematics. The molecular kinematics, while limited to a small region, shows clear disk like rotation around a PA$=137^\circ$ east of north. The molecular gas velocity dispersion is centrally concentrated with an average value of $\sigma\approx38$ km s$^{-1}$, in good agreement with previous works \citep{Brum2019}. The ionized gas maps seem to show a shifted rotation compared to the molecular gas, with a PA$=105^{\circ}$, and its velocity dispersion also present higher values along this axis. The higher dispersion values for the ionized gas could be produced by the jet interacting with the gas. This scenario agrees with the gas kinematics from Gemini/NIFS presented in \citet{Nandi2023}, with a jet piercing through the gas and exciting the gas along its direction as well as on cocoons around the jet. For NGC~4395 the JWST/NIRSpec data also allowed us to obtain kinematics for two coronal lines; [Si VI] at $1.96\,\mu {\rm m}$ and [Ar VI] at $4.52\,\mu {\rm m}$. The coronal line kinematics are similar to the ionized gas velocity rotation, with a east-west velocity gradient.

The \textbf{flux maps} show (Figure~\ref{fig:flux_maps}, bottom-row) a molecular and ionized gas distribution that is centrally concentrated, but also reveal a distinct blob of gas emission southwest of the nucleus. Compared to the Pa$\alpha$ and Br$\gamma$ lines, the molecular gas is more extended in the direction perpendicular to the radio jet and also along the blob. The coronal lines are also extended and show emission along the blob. This blob has been previously identified \citep{Denbrok15, Brum2019}, and could be the result of the accretion of a gas cloud or the ongoing accretion of a small gas-rich galaxy \citep{Brum2019}.
The H$_2$/Br$\gamma$ line ratio maps reveals AGN photoionization as the dominant excitation mechanism of the gas, while shocks associated with the jet or AGN outflows do not play an important role. However, shocks appear to contribute to the excitation east from the jet, on what could be clouds of gas being pushed away from the jet. Our results agree with the findings of \citet{Brum2019}, who found, using [FeII] emission, that AGN photoionization is the dominant excitation mechanism, but that shocks cannot be ruled out. We tested different binning strategies to discard the shocks line ratios could result from noise or bad S/N in bin edges. However, we continue to find this region to be shock dominated. Give the limited S/N of our data, we caution against a strong interpretation of the presence of shocks in NGC~4395. 

\section{Summary}
\label{sec:Summary}
We used JWST NIRSpec IFU spectroscopy for M87, M104, M94, M81, NGC~1052, NGC~4258, NGC~4395 from the ReveaLLAGN survey and the GTO data for NGC~5128 to extract stellar and gas kinematics in the inner $\sim100$ pc. The JWST/NIRSpec IFU data were heavily affected by sinusoidal undersampling noise or wiggles \citep{Law2023}. To properly extract the stellar and gas kinematics, we cleaned the spectra from wiggles using \texttt{WICKED} \citep{Dumont2025} in the nuclear $\sim0.5\arcsec$. We summarize the main finding of this work here:
\begin{itemize}
  \item The stellar kinematics reveal disk like structures for all the galaxies. The disk like rotation is accompanied with anti-correlated h$_3$ values as expected \citep{Krajnovic2008,Sande2017}. The stellar velocity dispersion maps for these objects show a rise to higher values towards the center associated with higher h$_4$ moment values as expected in the presence of a central supermassive black hole \citep[e.g][]{Cappellari2016}. Of the seven objects we study, the velocity dispersion for M94 and M104 stand out, showing more complex maps. M94's dispersion drops in the inner $0.3\arcsec$ to only increase in the very nuclear $0.05\arcsec$, which could be due to a counter-rotating kinematic structure. M104 shows a peanut-shape nuclear dispersion and an increase to higher dispersion away from the disk. The h$_3$ seems to also change sign in the inner $0.1\arcsec$. We speculate that this could be due to the presence of another kinematic structure in the very center of M104, like a nuclear star cluster. 
  \item The velocity maps for the ro-vibrational H$_2$ lines show disk like rotation. The rotation of the molecular gas is usually decoupled from the stellar disk, suggesting that the gas has been accreted and not settled with the stellar disk. The molecular velocity maps also show signs of non-rotation along the direction of the radio jet, for five of the eight LLAGN. This non-rotation is frequently associated with higher values of the velocity dispersion. 
  \item Our targets also show multiple hydrogen recombination lines from ionized gas, namely Pa$\alpha$, Br$\gamma$, Br$\beta$, and three transitions from the Pfund series. The S/N obtained with our JWST/NIRSpec data allowed us to construct velocity maps for the first two of these lines. The ionized hydrogen traced by Pa$\alpha$ and Br$\gamma$ reveals a disk like rotational structure comparable to that observed in the molecular H$_2$ lines, but it is less affected by the radio jet. However, the Pa$\alpha$ velocity dispersion attains higher values along the radio jet than those found for molecular hydrogen, and appears to effectively trace the jet’s impact.
  \item The emission of molecular hydrogen H$_2$ is usually more extended than the ionized hydrogen traced by Pa$\alpha$ and Br$\gamma$. In most cases, the Pa$\alpha$ and Br$\gamma$ emission shows a steeper flux gradient, whereas the molecular hydrogen generally exhibits tail-like features aligned with the radio jet. In two of our lowest Eddington ratio sources, M87 and M94, we detect off center molecular hydrogen structures that are not clearly associated with either the radio jet or star-forming regions. The complementary MIRI observations of M94 presented by \citet{Kameron2025} also show the gas blob west of the nucleus. In M87 the MIRI observations do not have show molecular hydrogen lines emission (see Figure 1 \citet{Kameron2025}), and we cannot determine whether a similar elongated structure is present. However, \citet{Kameron2025} found that the line profiles in M87 are double-peaked, suggesting the presence of an additional kinematic structure embedded in the nuclear region
  \item The measured ratio between molecular and ionized hydrogen emission, traced by H$_2$ 1-0 S(1) and Br$\gamma$, suggest that AGN photoionization is the primary ionizing source in the nuclear regions of all LLAGN in our sample, regardless of their Eddington ratios. Based on this line ratios we do not find evidence of strong star formation in the nuclear regions of our target galaxies. While we find visual evidence of the molecular gas PA being aligned with the radio jet, suggesting its kinematics may be affected by it, the line ratios suggest that shock ionization, from outflows or the jet, may only play a minor role. Shocks become the dominant ionization mechanism only in localized areas of NGC~5128 and possibly also in NGC~4395 and NGC~1052. For NGC~5128, the shocks are spatially associated with the radio jet and also with gas ring structures. NGC~1052 seems to show small regions of shock ionization along the jet,
  while in NGC~4395, they are found away from the jet axis.
  However, the limited S/N of the Br$\gamma$ line made the line ratio maps noisy impacting our ability to confidently characterize the dominant ionization mechanism, specially in cases like M87 and Sombrero, where large parts of the field-of-view needed to be masked out.
  \item Coronal lines such as [Si VI], [Mg VIII] and [Ar VI] are detected in the nuclear aperture spectra in the ReveLLAGN sample. However, due to their low S/N at the individual spaxel level, even after binning the data, kinematic maps were only attainable for our highest Eddington ratio objects, NGC~5128 and NGC~4395. In NGC~5128 coronal line emission extends along the direction of the radio jet and displays non-rotational motion. NGC~4395's coronal lines show similar rotational motion to the ionized hydrogen gas. We recommend longer exposure times for the study of NIR coronal line kinematics in the future. 
  \item Overall, we detect high ionization coronal lines in all LLAGN when using nuclear aperture extractions. The main distinction we observe among the LLAGN in our sample is not whether coronal lines are present, but how strong they are. In NGC~5128 and NGC~4395, our highest Eddington ratios objects, the coronal lines are bright enough that we can to obtain their spatially resolved kinematics. While for the rest of the sample with lower Eddington ratio this is not possible. In NGC~5128, the coronal lines is spatially and kinematically align with the direction of the radio jet. The evidence for shock excitation along NGC~5128 jet direction, suggest the jet may be responsible for the extended coronal emission. In NGC~4395, in contrast, the coronal line emission show very similar kinematics as the recombination lines.
  
\end{itemize}

The gas kinematic maps presented in here provide the basis for future work on gas dynamical modeling of the supermassive black hole masses in this sample.

\section*{Acknowledgement}
LCH was supported by the National Science Foundation of China (12233001) and the China Manned Space Program (CMS-CSST-2025-A09)

\bibliography{References}
\bibliographystyle{aa}

\end{document}